**ISTANBUL TECHNICAL UNIVERSITY**
**FACULTY OF COMPUTER AND INFORMATICS**

# A STUDY ON TRENDS IN INFORMATION TECHNOLOGIES USING BIG DATA ANALYTICS

**Graduation Project**

**Mahmut Ali ÖZKURAN**
**040000815**

**Department: Computer Engineering**

**Advisor: Asst. Prof. Dr. Ayşe Tosun Mısırlı**

This Page Intentionally Left Blank

**ISTANBUL TECHNICAL UNIVERSITY**
**FACULTY OF COMPUTER AND INFORMATICS**

# A STUDY ON TRENDS IN INFORMATION TECHNOLOGIES USING BIG DATA ANALYTICS

**Graduation Project**

**Mahmut Ali ÖZKURAN**
**040000815**

**Department: Computer Engineering**

**Advisor: Asst. Prof. Dr. Ayşe Tosun Mısırlı**

This Page Intentionally Left Blank

## Özgünlük Bildirisi

1. Bu çalışmada, başka kaynaklardan yapılan tüm alıntıların, ilgili kaynaklar referans gösterilerek açıkça belirtildiğini,
2. Alıntılar dışındaki bölümlerin, özellikle projenin ana konusunu oluşturan teorik çalışmaların ve yazılım/donanımın benim tarafımdan yapıldığını bildiririm.

İstanbul, 29.05.2015

Mahmut Ali ÖZKURAN

This Page Intentionally Left Blank

# A STUDY ON TRENDS IN INFORMATION TECHNOLOGIES USING BIG DATA ANALYTICS

# (SUMMARY)

We are living in an information era from Twitter [1] to Fitocracy [2]; every episode of peoples' life is converted to numbers. That abundance of data is also available in information technologies. From Stackoverflow [3] to GitHub [4] many big data sources are available about trends in Information Technologies.

The aim of this research is studying information technology trends and compiling useful information about those technologies using big data sources mentioned above. Those collected information might be helpful for decision makers or information technology professionals to decide where to invest their time and money.

In this research we have mined and analyzed StackExchange and GitHub data for creating meaningful predictions about information technologies. Initially StackExchange and GitHub data were imported into local data repositories. After the data is imported, cleaning and preprocessing techniques like tokenization, stemming and dimensionality reduction are applied to data.

After preprocessing and cleaning keywords, their relations are extracted from data. Using those keywords data, four main knowledge areas and their variations, i.e., 20 Programming Languages, 8 Database Applications, 4 Cloud Services and 3 Mobile Operating Systems, are selected for analysis of their trends.

After the keywords are selected, extracted patterns are used for cluster analysis in Gephi [5]. Produced graphs are used for the exploratory analysis of the programming languages data. After exploratory analysis, time series of usage are created for selected keywords. Those times series are used as training and testing data for forecasts created using R's "forecast" library.

After making forecasts, their accuracy are tested using Mean Magnitude of Relative Error and Median Magnitude of Relative Error.

This Page Intentionally Left Blank

# A STUDY ON TRENDS IN INFORMATION TECHNOLOGIES USING BIG DATA ANALYTICS

# (ÖZET)

Günümüz dünyasında kredi kartıyla yapılan alışverişten toplu taşıma kullanımına kadar yaşamımızın her noktasında veri oluşturuyoruz. Bu oluşturulan büyük veri şehir planlamacılığından iş zekâsına, sigortacılıktan reklamcılığa hayatımıza birçok noktada etki etmektedir. Trafik işaretlerinin kullanımı, kişi bazında indirimli ürünler, kişiye uygun detaylı risk analiziyle daha uygun sigorta poliçeleri büyük verinin hayatımıza etkisinin sadece küçük örnekleridir.

Büyük Veri bu bahsettiğimiz kaynakların yanı sıra Bilgi Teknolojileri konusunda da pek çok imkânlar sunmaktadır. Özellikle yapılacak yatırımların uzun ömürlü ve yüksek getirili olması için yatırım yapılacak teknolojinin doğru seçilmesi gerekmektedir. Gartner'e [6] göre dünya genelinde bir senede Bilgi Teknolojilerine harcanan para 3,8 trilyon Amerikan dolarından fazladır. Bu pazarın büyüklüğü yapılacak her tercihte birçok seçeneğin göz önüne alınmasını gerektirmektedir. Tercihlerin çokluğu ve pazarın büyüklüğü göz önüne alındığında her bir tercihin yapılmasında mümkün olan her türlü yardım çok önemli duruma getirmektedir. İşte bu durumda yapılacak yatırımı en az riskli hale getirmek için geçmiş veriler kullanılarak modeller oluşturmak ve öngörülerde bulunmak en çok kullanılan yöntemlerden biridir.

Bu çalışmanın amacı yukarıda bahsi geçen veri kaynaklarını kullanarak anlamlı bir bilgi kaynağı oluşturmaktır. Bu oluşturulacak bilgi kaynağının, bilgi teknolojileri konusunda yatırım yapacak kişilere ve bilgi teknolojileri konusunda çalışan profesyonellere hangi teknolojiye zaman ve para yatırmaları konusunda karar verirken yardım etmesini ummaktayız.

İşte bu koşullar altında öncelikle araştırma yapılacak "Bilgi Alanları" IEEE'nin yayınladığı Curriculum Guidelines for Undergraduate Degree Programs in Computer Engineering [7] yardımıyla tespit edildi. Bahsi geçen yayında bulunan bilgi alanları içinden günümüzün önde gelen bilgi teknolojileri göz önüne alınarak Programlama Dilleri, Taşınabilir Cihazlar İçin Geliştirilmiş İşletim Sistemleri, Bulut Hizmetleri ve Veritabanı Teknolojileri seçildi.

Bilgi Alanlarımızı tespit ettikten sonra her bir Bilgi Alanının önde gelen teknolojileri işlenmemiş verimizin analiziyle en çok bahsi geçen teknolojiler olarak seçildi.

Öngörü modeli oluştururken öngörünün kalitesini etkileyen en önemli girdiler verilerin çokluğu ve doğruluğudur. İşte bu nokta da Büyük Verinin öngörü konusundaki önemi ortaya çıkmaktadır. Çünkü Büyük Veri teknolojileri kullanarak oluşturacağımız bir Veri Ambarı sayesinde kolayca ulaşabileceğimiz bol ve doğru veri sayesinde daha güvenilebilir öngörüler oluşturabiliriz.

Bu Veri Ambarını oluşturmak için ilk adımımız güvenilir bir veri kaynağı bulmak olmalıdır. Bilgi Teknolojileri konusunda pek çok veri kaynağı bulunmaktadır. Bu veri kaynaklarının seçiminde kaynağın tarafsız olması ve erişimin serbest olması en önemli noktalardandır. Ayrıca daha önce bahsedildiği üzere veri kaynağı seçilirken verinin çokluğu da öngörü oluşturmanın gereksinimlerinden biri olarak göz önüne alınmalıdır. Bu ölçütler göz önüne alındığında StackExchange ve GitHub iki uygun seçenek olarak karşımıza çıkmaktadır.

StackExchange 4,5 milyondan fazla kullanıcısıyla internetin en büyük soru-cevap topluluğu. StackExchange gerek kullanıcıları tarafından yönetilmesi olsun gerek her konu için ayrı topluluklara sahip olmasıyla olsun üst kalitede bir veri yapısına sahip bir oluşum. StackExchange in sahip olduğu alt-topluluklardan en dikkat çekici olanı yazılım geliştirme konusunda uzmanlaşmış olan Stack Overflow topluluğu. Stack Overflow'da yazılım geliştirme konusunda karşılaşılan her türlü probleme konudaki uzmanlar tarafından cevap verilmesi çok olası.

Stack Exchange verisi özellikle organik olarak büyümesi yani bu durumda herhangi bir teknolojiyi kullanırken karşılaşılan problemlere yanıt ararken oluşması nedeniyle gerçekçi bir veridir. Reklam amaçlı farkındalık artışı veya anketlerde karşılaşılan yönlendirmeler sonucu gerçek fikrin veride temsil edilmemekte olması gibi problemler bu veride minimum seviyededir.

StackExchange verisinin yapısal durumuna baktığımızda verinin en önemli yapının "Post" ve "Tag" olduğunu görüyoruz. "Post" soru ve cevap girdilerinin kayıtlarının tutulduğu, "Tag" ise bu soru ve cevap girdilerinin sınıflandırılmasında kullanılan anahtar kelimelerin tutulduğu veri yapısı olarak tasarımlanmış. "Tag" veri yapısı ve "Post" veri yapısının içeriğindeki metin girdi olan "Body" alanı sayesinde Soru ve cevap girdilerinin sınıflandırılması mümkün hale gelmekte. Özellikle "Body" alanının içeriğinin kelimler halinde parçalanması ve bu kelimelerin köklerine ayrılması sayesinde anahtar kelimeler daha efektif şekilde sınıflandırılma yapılmasını sağlayabilecektir.

GitHub web tabanlı bir sürüm kontrol sistemidir. Serbest kaynak kodlu bir yazılımların kaynak kodu bu sitede tutulmaktadır. Linux çekirdeğinin ve Microsoft .NET kütüphanesinin kaynak kodu bu sitede barındırılmaktadır. Verisinin büyümesi içeriği hakkında StackExchange verisiyle benzer şekilde yorumlarda bulunulabilir. Veri içeriğinin çoğalması ihtiyaç halinde gerçekleşmektedir. Yani bir yazılım gelişmesi veya bir yazılımda ki hatanın düzeltilmesi GitHub'ın veri setine ilave girdi olarak eklenmektedir.

GitHub'ın verisi "Event" yani Hadiselerden oluşmaktadır [8]. Hadise olarak bahsedilen olayların içeriğindeki "repository_language" özelliği bu girdinin sınıflandırılması açısından önemli bir niteliğe haizdir. Bu alan sayesinde o dilin belli bir zaman aralığında ki kullanılma miktarını tespit edebiliriz. Bu özelliği kullanarak oluşturan zaman serileri daha sonra öngörü oluştururken kullanılabilecektir.

Bu iki veri kaynağı kullanarak bir veri ambarı oluşturulmasına karar verildikten sonra veri alımı için StackExchange veri dökümü [9] ve GitHub veri dökümü [10] kullanıldı. Her iki veri dökümü de geliştirilen bir uygulama [11] sayesinde yerel MongoDB veri tabanına yüklendi.

Veri tabanında bulunan verinin kelimler halinde parçalanması ve bu kelimelerin köklerine ayrılması işlemi yapılmış ve elde edilen anahtar kelimelerle bu verilerin sınıflandırılması sağlanmıştır. Ardından bu anahtar kelimelerin kullanılmaları zaman aralığına göre ölçülerek ay, 3 ay, 6 ay ve yıllık aralıklarla StackExchange ve GitHub verileri için ayrı ayrı zaman serileri oluşturulmuştur.

Zaman serilerinden ayrı olmak üzere anahtar kelimelerin birbiriyle ilişkileri göz önüne alınarak yıllık tabanda anahtar kelime çizgesinin verisi oluşturulmuştur. Bu oluşturulan veri Gephi [5] uygulaması kullanılarak incelenmiş ve üzerinde keşifsel veri analizi yapılmıştır. Oluşturulan çizge verisinin metriklerinin yorumlanması ise daha ileriki bir araştırmaya bırakılmıştır.

Zaman serileri ile öngörü oluşturmak için R dilinin “forecast” kütüphanesi [12] kullanılmıştır. Bu kütüphanenin en büyük özelliği verinin tipine göre öngörü algoritmasını değiştirebilmesidir. Bu araştırmada tutarlı sonuçlar elde etmek amacıyla “ets” yani üstel yumuşatma algoritması kullanılmıştır. Bu metodun tercih edilmesinin sebebi en son veri noktasına en yüksek katsayı çarpanını atarken daha eski veri noktalarına üstel olarak azalan büyüklükte çarpanları katsayı olarak atamasıdır. Burada tek olumsuz nokta gerçekçi bir öngörü için öğretici veri noktası miktarının fazlaca olması gerekmesidir. Gerekli veri noktasının fazla olması gerekliliği ile yıllık ve 6 aylık öngörülerde problem yaratmış ve bu zaman aralılarının analizi araştırmadan çıkarılmıştır.

3 aylı ve aylık öngörüler oluşturulduktan sonra bu öngörülerin deneme verisi ile uyumları tespit edilmiş ve her bir öngörünün Ortalama Bağıl Hata (MRE) ve Medyan Bağıl Hata (MdRE) değerleri hesaplanmıştır. Bu değerler tahminlerin doğruluğunun analizinde kullanılmıştır. Seçilen teknolojinin gelecekte nasıl bir yol izleyeceğinde ise sapmanın büyüklüğü ve öngörünün hangi tarafında (az veya çok) bulunduğu göz önüne alınmıştır.

This Page Intentionally Left Blank

# CONTENTS

# 1 INTRODUCTION

According to Gartner [6] global Information Technologies Market is bigger than $3.8 trillion. This huge size of the market causes decision making increasingly more difficult. Today there is no single best solution for any problem in Information Technologies. While there are many decision making tools and frameworks available; every decision have been made needs foundation based on data. Decision making needs wisdom and wisdom needs process of data to become one [13].

Knowledge Hierarchy [13], suggests 'Knowledge Pyramid' (Figure 1) which points out road to Wisdom from Data is going through Information and Knowledge. To reach Wisdom one should follow the path through Data, Information and Knowledge.

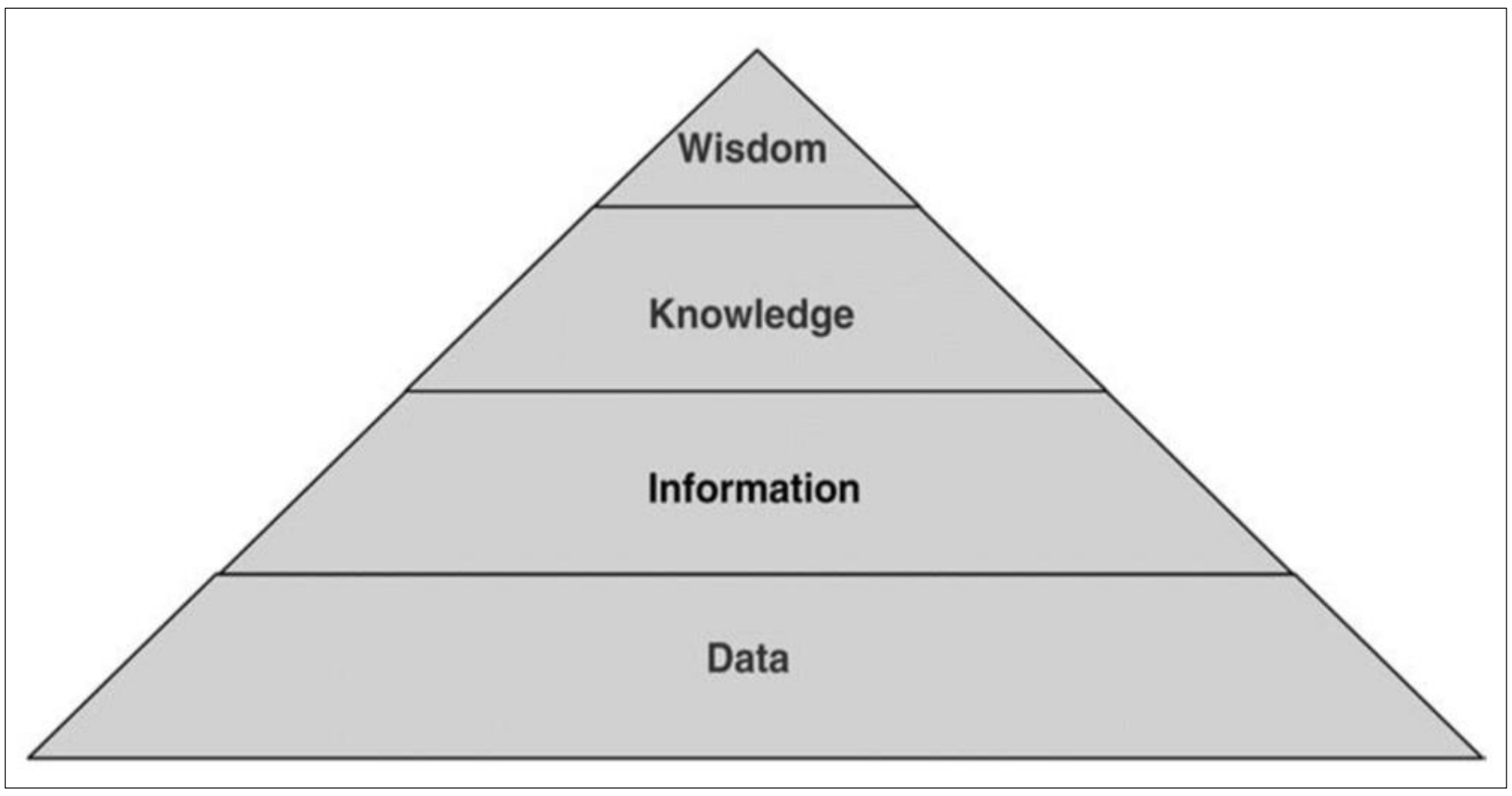

***Figure 1.1:*** *Knowledge Pyramid*

Data is raw acquisition of differences in physical states around us [14]. As result of acknowledging every difference in states as Data is resulting huge amount in volume. This huge volume is useless until relations and other transformation in content made in this Data. This transformed type of Data is described as Information. For example Data is the list of sales in a supermarket, Information is result of analysis in which percent something buy with another.

Knowledge is result of one step further analysis of Information. For example if we have seen many time beer with diapers bought together we can say that if someone buy diapers he will probably buy beers too. This deduction from given Information is called Knowledge.

Wisdom is more like become aware of situation in the Information. Cognitive filtering of Information leads us to Wisdom. If we make addition to previous example, putting diapers and beers next to each other to rise sales, is result of the Wisdom.

Today still differences are present between Wisdoms' of human intelligence and computational power. This point is the only transition in Knowledge Pyramid which still needs human cognitive ability. Rest of this transitions could be achievable with computational methods.

In this study our target is creating Knowledge for Wisdom which is giving decisions about future trends of Information Technologies.

As our study topic is very broad, many research studies have been conducted on this topic. As information technology trends is a hot topic, researchers focus on smaller knowledge areas.

An earlier study [15] made by Doukidis, Smithson and Lybereas in 1994 is a good example on trends in Information Technology research. The study was made from a management perspective on the information technologies and small businesses. The study is based on conducted surveys made with the companies' decision makers on information technologies. Authors present interesting results such as lowered weight of the hardware suppliers in outside advice or changing structure of companies with computers [15]. Results highlight the impacts of information technology trends in small businesses. Changes in information trends have become the second degree in the results. Our research is different than this research in the meaning of both scope (as our scope is Information Technology Trends) and data collection methodology.

Another research was conducted by Rosen and Shihab [16]. Authors mined StackOverflow data in scope of mobile device development and general issues faced related to mobile development. Data collection methodology of their research includes StackOverflow data dump which is similar to our methodology. But it's more detailed in mobile development in nature, it does not concentrate on trends of information technology in general, but it focusses more on questions of mobile application developers.

Our research is also conducted on StackExchange and GitHub which are both big data sources on information technology, especially on application development. Both data sources has very large datasets produced by millions of individuals. Only StackOverflow site which is a sub community in StackExchange network has more than 4,250,000 users by the May of 2015.

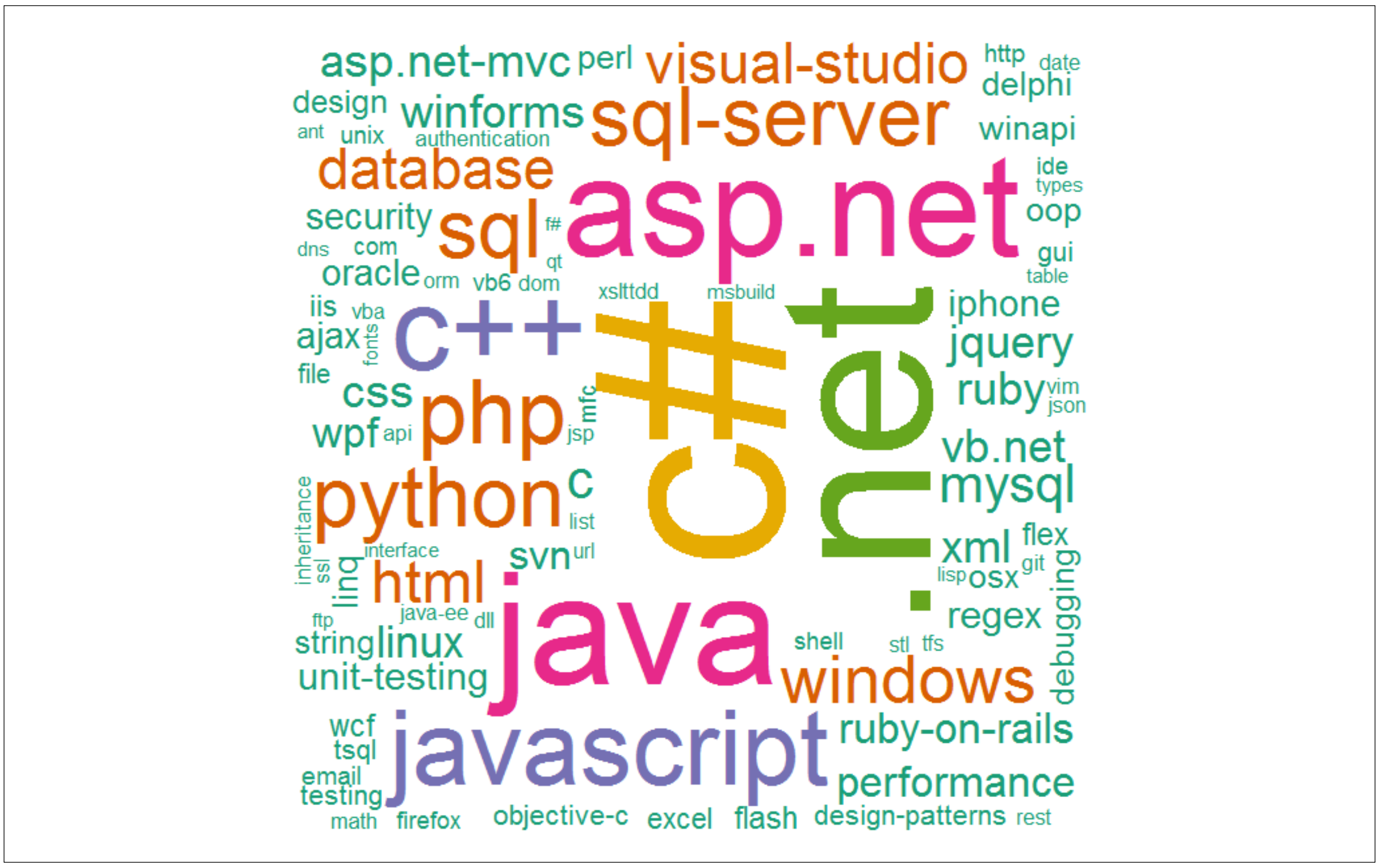

***Figure 1.2:*** *Word Cloud of Stackoverflow Posts - 2008*

StackExchange is the biggest Question and Answer platform in the world with more than 25 million posts in different knowledge areas. StackExchange is a perfect data set for our study as it is not just data consists of text also classified by tags. Figure 1.2 shows an example word cloud that created from those tags. This dataset is publicly available as a XML file which contains all data from August 2008 to March 2015 from archive.org web site [9].

GitHub is the biggest web-based Git repository hosting service. It hosts great software projects like Linux and .NET library source code. As an industry standard in open source projects huge open source projects like Bootstrap [17], Node.js [18], jQuery [19] and Ruby on Rails [20] are hosted in GitHub. Also with more than 9 million contributors and more than 100,000 activities like bug fix, pull requests or repository creation has been happening every day. The dataset is publicly available freely at *data.githubarchive.org* in JSON format starting from 02/12/2011. As pulled data is limited to start of year 2015 its size is nearly 25 GBs.

This Page Intentionally Left Blank

# 2 PROJECT DESCRIPTION AND PLAN

The aim of this research is studying information technology trends and compiling useful information about those technologies using big data sources mentioned above. Those collected information might be helpful for decision makers or information technology professionals to decide where to invest their time and money. As we mentioned before, wisdom needs knowledge and knowledge must be collected from information which is filtered from correct data.

As our research is about studying Information Technologies trends, we first have to define the Knowledge Areas that we are going to study. To decide the Knowledge Areas we have used Curriculum Guidelines for Undergraduate Degree Programs in Computer Engineering by IEEE [7]. We have selected Programming Languages, Mobile Operating Systems, Database Systems and Cloud Services among 18 Knowledge Area defined in this book.

Under these 4 Knowledge Areas we have selected key technologies using most active keywords in our raw data. Based on the most used keyword we have selected key technologies that we will use in our research.

Project duration is defined as 170 days. It starts on October 2014 and finishes on May 2015. Project is planned to complete in 4 stages. Planned stages are Preparation of Data Collection Infrastructure, Data Collection, Data Analysis and Reporting. Table 2.1 displays Work Breakdown Structure of the project.

| WBS | Task Name |
|---|---|
| **1** | **A study on trends in information technologies using Big Data Analytics** |
| **1.1** | **Preparation of Data Collection Infrastructure** |
| 1.1.1 | Software Installation |
| 1.1.2 | Analyzing Required Data Structure |
| 1.1.3 | Creation of Databases |
| **1.1.4** | **Testing of Data Infrastructure** |
| 1.4.1 | Implementation of StackExchange Data Connector |
| 1.4.2 | Collecting Test Data |
| 1.1.4.3 | Fixing Bugs |
| **1.2** | **Data Collection** |
| 1.2.1 | Implementation of Data Connectors |
| 1.2.2 | Data Collection |
| **1.3** | **Data Analysis** |
| 1.3.1 | Software Installation |
| 1.3.2 | Connection of Data With Analysis Software |
| 1.3.3 | Preliminary Data Analysis |
| 1.3.4 | Improvements on Data Analysis |
| 1.3.5 | Complete Data Analysis |
| 1.4 | Reporting |

**Table 2.1:** Work Breakdown Structure of Project

Preparation of Data Collection Infrastructure is initialization stage of the project. Software infrastructure is prepared in this stage. Two virtual computers created and installed using Microsoft Windows Server 2012 and Ubuntu was installed as operating system. Microsoft SQL Server 2012 was installed into Windows Server machine and MongoDB 2.6 was installed on Ubuntu machine. SQL Server is used to store structured data and MongoDB for document based data.

In Data Collection stage StackExchange and GitHub data was imported to the above mentioned databases. Due to the nature of the data, both datasets were first imported into MongoDB as documents. An application in C# was developed for import process.

In Data Analysis stage data mining methods like Tokenization and Stemming were applied onto data using developed software. Keyword analysis and data classification was made and text dataset was converted into time series and graph data sets.

# 3 THEORETICAL KNOWLEDGE

## 1. DATA CLEANING & PREPROCESSING

We have applied some of standard data mining methods onto our raw data. Purpose of implementing those methods is better performance in means of both time and accuracy. If we weren't applied those methods on to raw data precision of our results would be suffer.

### 1.1. TOKENIZATION

Analysis of text data is requires conversion of the big chunk of data to smaller, more meaningful pieces. Tokenization [21] is the starting point for this process as it breaks sentences into words. Implementation of tokenization depends on the purpose of the data analysis. We may remove only white spaces or also we can remove punctuation while tokenizing the text.

### 1.2. STOP WORDS

Stop words are the words that have importance in the grammar or creation of an idea but that does not have importance in themselves. For example; while words like "I", "am" and "the" are the common words that used in presentation of any idea they are so common not have any meaning for data mining.

Removing stop words from text data is the first step on many text search/classification algorithms. For example MySQL have a 550 word English stop words that are used in text mining for better performance [22].

### 1.3. STEMMING

Stemming is a technique that used for merging words with the same root. For example; "radiation", "radioactive" and "radiologist" words have same root "radio". This function gives the chance of finding related documents even given word is not included in the results. As there many stemming algorithms available, we use Porter2 stemming algorithm in this research [23].

### 1.4. DIMENSIONALITY REDUCTION

Dimensionality reduction [21] is a technique that is used in machine learning and statistics to reduce number of variables for easier understanding of data. Dimensionality reduction is not only important by the means of easing the understanding of data but it is also important for removing the physical limitations like time and computing power. Especially in huge data sets like ours it is imperative to use dimensionality reduction as memory bottleneck became biggest limitation of the research. In our research we used simple exploratory data analysis techniques to decide our critical points in the data and reduced dimensionality of our data using this results.

# 2. DATA MINING

## 2.1. PATTERN DISCOVERY

Patterns are set of items, structures that occur together in a dataset. Patterns are important properties of the datasets as they are inherent regularities in a dataset [24]. Discovering those inherent regularities in the data set is essential in many data mining tasks. For example many Business Intelligence methods like Market Basket Analysis and Sale Campaign Analysis are arises from simple pattern discovery methods. In our research we have used graphs to visualize relations in our data.

# 3. DATA ANALYSIS

## 3.1. FORECASTING

Forecasting [25] is making predictions about a future data value. It is helping to cope with uncertainty of the future. Forecasting has a usage in every area that needs prediction about the future. Especially finance sector is relying on forecasting heavily. In our research forecasting used for predict the next values of the time series.

## 3.2. EXPONENTIAL SMOOTHING

Exponential smoothing [12] is a statistical method that used in forecasting. The method is uses previous values to predict what will be the next value become. It is similar to moving average except Exponential Smoothing uses recent values with a bigger coefficient in calculation. In Exponential Smoothing as data become older the coefficient of the value is decreasing exponentially.

In our research we have used the implementation of Exponential Smoothing (ETS) in R’s “forecast library.

## 3.3. CLUSTER ANALYSIS

Cluster analysis [21] is grouping objects in a dataset based on their similarity. For example grouping product related to their ingredients can be considered as clustering. In our research keyword data used for clustering of data. Count of the keyword appearances and relations between those appearances are used for cluster analysis

# 4 ANALYSIS AND MODELLING

Desired output of this research is creating a trend model for well-known information technologies. As information technologies are the core of the modern life there are many parameters that affects the trends of information technologies. A new trend in young generations like online video sharing may lead unnoticed new technology requirements. To follow up and guessing what will be next big thing is plausible but hard. As we mentioned before especially in general technology trends so many things can affect what will be the next "trendy" technology. To find what will be trend next we should follow many data sources from patents to newspapers.

Patents are the first class sources that we use what will be next technology. They have been used at least until last few decades. As patent trolling grew up and became a new industry many firms started to earn more money with lawsuits than their products. Patent trolling also causes many meaningless patents to be taken. Insignificant amount of patents in a huge pile are not meaningful for a statistical research.

For a better trend creation we focus on professionals of the Information Technologies sector. StackExchange is become our first data source to look at. It is a huge Q&A community with millions of information technology professionals. As it is unbiased and result orients community, it is an ideal data source for trend analysis.

Another good source for analysis of technology trends is GitHub. As it is only serves for open source community it is still good data source for a huge part of the Programming Languages. Today nearly all kind of applications have an open source counterpart. From database servers to operating systems there is at least one open source application present that widely used in the market. Especially in last decade most of software companies have decided to earn most of their income from services not software licenses. Success of this model caused many companies start creating open source solutions. Today Microsoft made his .NET library open source, Oracle develops open source database systems as MySQL. Those are all effects of the new market model. And if we want to include this conjuncture to our research we should add GitHub data to our dataset.

To create a model first thing we should do creating a dataset to make our models on. StackExchange and GitHub selected as data sources to generate dataset. StackExchange have multiple Question and Answer communities that specialized in different topics. Stack Overflow, Mathematics and Super User are some of the biggest communities of the StackExchange.

To create the data that supporting our model we should define detectable points in data we may use in our research. In our research we have analyzed our raw data and detected keywords which are related to our already decided Knowledge Areas. "*Keyword*" is distinctive words or texts that are belong to our interested technology.

Data structure of the StackExchange is same for all communities. Two main data structure are "*User*" and "*Post*". Additionally "Post" has an external data structure name as "*Tag*" for classification. "*User*" structure is containing information about the users like "*Age*", "*Email Address*" and "*Location*". As we researching Information Technologies trends those information is not so meaningful and related to our research. But in “*Post*” and “*Tag*” data structure story is a little difference. “*Post*” data structure have important text based data “*Body*”. “*Body*” has all the text data of the “*Post*” and a good source for text mining. Another important structure of StackExchange dataset is “*Tag*”. “*Tag*” is important because they are used to classify the posts. Especially as those tags are moderated thru the community they are very accurate. Even if a “*Posts*” “*Body*” did not include the “*Keywords*” that related to its content “*Tags*” are useful to classify this post.

Stack Overflow is a Question and Answer community for programmers. With more than 25 million posts about everything programmable it’s a huge dataset as a whole. A sample Question may be found below.

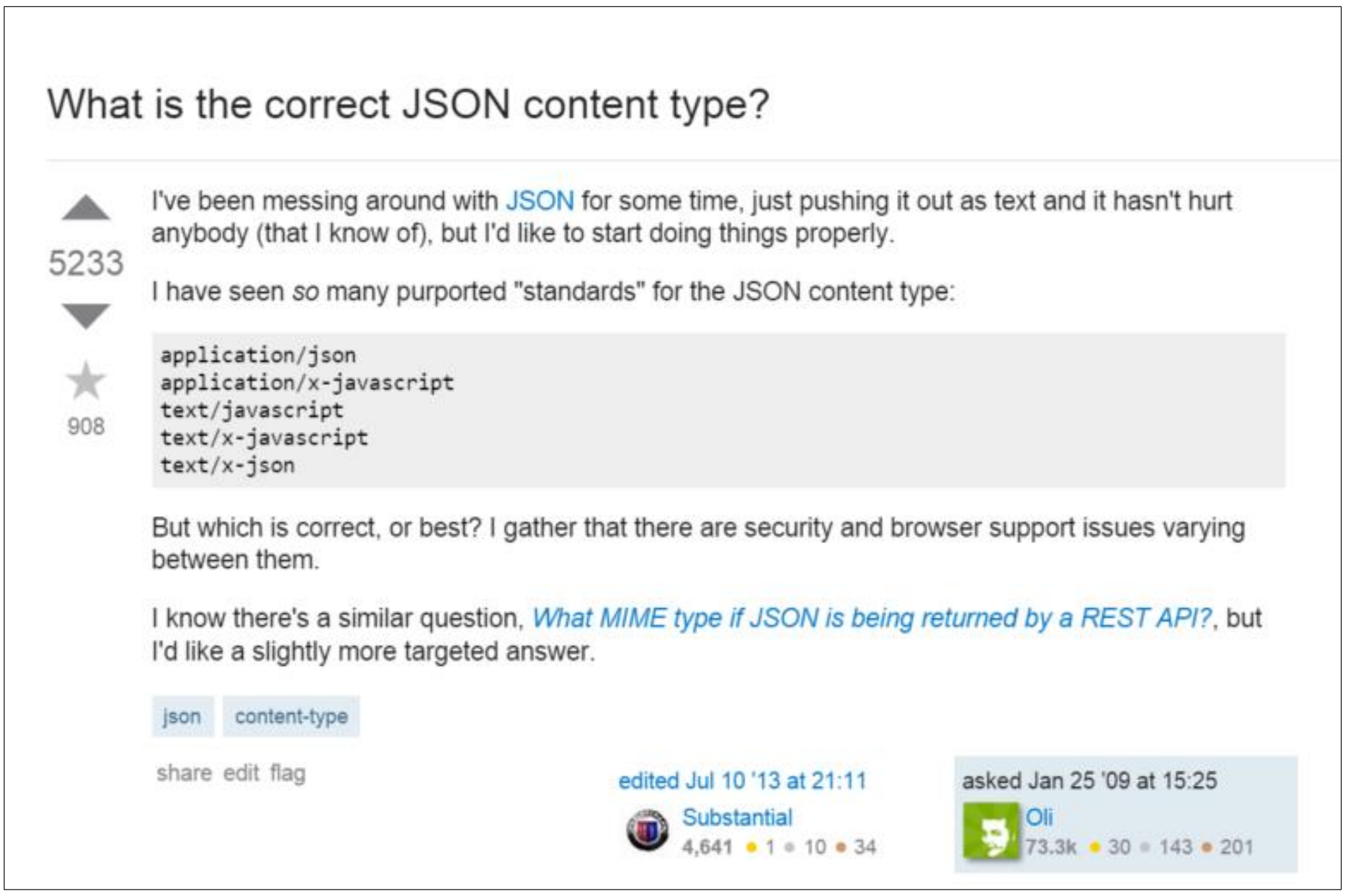

**Figure 4.1:** Sample Stack Overflow Post

Questions that posted on StackOverflow are mostly about the problems that arise when developing an application. At this context those questions could be good sources of information about programming language trends. Both classification of the questions with tags and text of the question will give us meaningful information related to our research. Text data is processed using data mining techniques mentioned above and combined with tag data to classify posts.

Database Administrators community in StackExchange is also taken into account as it have useful information about databases. As Database Administrators data structure is same as Stack Overflow data structure same methods applied to Database Administrators community data.

GitHub is second source of our researches data set. GitHub data consists of JSON documents which are consists of properties of the events created by users’ actions. There are currently 25 events presented [8]. The document of the each event differs from one another. Even all documents have different properties every event document has a property called “repository_language” which allows us to classify the document of the event.  Those events are listed in Appendix B.

This Page Intentionally Left Blank

# 5 DESIGN, IMPLEMENTATION AND TEST

## 1. DESIGN & IMPLEMENTATION

The data mentioned in the modelling section are downloaded and imported into database servers as documents. Importing process carried out using developed applications. After completion of importing data normalization cleaning methods like tokenization, removing stop words and stemming applied on to data. The numerical results of those applications are summarized in Data Summary section.

Cleaned data are analyzed and extracted keywords to classify data. A keyword in text data is compared with the data in the "*Tag*" of a "*Posts*" in StackExchange and the "*repository_language*" in a "*Event*" of GitHub. After the comparison "*Keyword*" and "*Synonym*" combined as the "*Tag*".

Using this "*Tag*" data, "*Posts*" and "*Events*" are classified with groups of Tags. This classification exported as time series and graph relation data. Classifications algorithms are combined into a library and placed onto NuGet for public usage [11].

Then, analysis of time series metrics are made by R language. The "*forecast*" library [26] of R have many useful functions for forecasting. In this research ETS method of this library is used for forecasting. The biggest benefit of this forecasting function is using different exponential smoothing methods for forecast depending on the training data. In the disadvantage point this forecasting method is if there is not enough data points it predicts only next data point. Especially lack of data points in GitHub data is effected this forecast method. To overcome this problem we have predict every single prediction point with new forecast.

Another data analysis method used in this research is graphs. Relation graphs produced and drawn but metric analysis of those graphs are left for a future research. Relation graph data is created by the developed application mentioned before. Created data format is "*gdf*" which includes nodes, edges and weights of both. Gephi [5] application used in graph creation stage.

## 2. DATA SUMMARY

Some numbers related to our data is listed below.

| | |
|---|---|
| Size of Stack Overflow Post Data (as XML) | 34,647,001,063 bytes |
| Size of Stack Overflow Post Data (as MongoDB JSON) | 45,795,208,928 bytes |
| Size of GitHub Data (as SQL) | 26,293,878,411 bytes |
| Size of Database Administrators Post Data (as XML) | 105,293,535 bytes |
| Total Posts in Stack Overflow Data | 24,120,523 |
| Total Posts in Database Administrators Data | 76,490 |
| Total Events in GitHub Data | 231,412,947 |
| Token count of combined data after tokenization | 112,168,012 |
| Unique Token count of combined data | 68,512,226 |
| Unique Token count of combined data Alpha only | 13,824,601 |
| Unique Token count of combined data Alpha and Lowercase | 11,064,160 |
| Unique Tag Count in StackExchange | 40,632 |

## 2.1. STACKOVERFLOW

As seen as in Table 5.1 and Table 5.2 StackOverflow has more than 600,000 posts made by users which equals more than 20,000 questions and answers every day.

**Stackoverflow Posts Per Month**

| | 2008 | 2009 | 2010 | 2011 | 2012 | 2013 | 2014 | 2015 |
|---|---|---|---|---|---|---|---|---|
| Jan | | 47,342 | 131,121 | 238,693 | 353,834 | 480,900 | 587,856 | 581,906 |
| Feb | | 52,319 | 133,169 | 247,154 | 375,358 | 468,342 | 581,464 | 591,816 |
| Mar | | 60,093 | 153,342 | 299,752 | 404,873 | 530,822 | 643,012 | 671,983 |
| Apr | | 62,692 | 147,990 | 286,142 | 392,252 | 525,342 | 601,434 | 687,003 |
| May | | 76,245 | 153,242 | 301,382 | 409,017 | 506,359 | 595,982 | |
| Jun | | 82,929 | 162,203 | 301,110 | 397,593 | 481,613 | 569,866 | |
| Jul | 14 | 94,991 | 176,250 | 307,072 | 432,751 | 536,234 | 620,321 | |
| Aug | 11,511 | 95,233 | 183,998 | 321,813 | 430,360 | 516,657 | 572,211 | |
| Sep | 42,342 | 96,408 | 178,340 | 305,788 | 402,928 | 509,215 | 583,180 | |
| Oct | 44,113 | 106,298 | 185,809 | 305,104 | 453,995 | 568,495 | 608,314 | |
| Nov | 37,960 | 111,977 | 204,031 | 329,786 | 449,700 | 545,600 | 588,040 | |
| Dec | 35,942 | 109,674 | 204,242 | 313,395 | 411,635 | 512,359 | 540,665 | |

**Table 5.1:** Stack Overflow Posts Per Month

**Stackoverflow Posts Per Day (Monthly)**

| | 2008 | 2009 | 2010 | 2011 | 2012 | 2013 | 2014 | 2015 |
|---|---|---|---|---|---|---|---|---|
| Jan | 0 | 1,527 | 4,230 | 7,700 | 11,414 | 15,513 | 18,963 | 18,771 |
| Feb | 0 | 1,869 | 4,756 | 8,827 | 12,943 | 16,727 | 20,767 | 21,136 |
| Mar | 0 | 1,938 | 4,947 | 9,669 | 13,060 | 17,123 | 20,742 | 21,677 |
| Apr | 0 | 2,090 | 4,933 | 9,538 | 13,075 | 17,511 | 20,048 | 22,900 |
| May | 0 | 2,460 | 4,943 | 9,722 | 13,194 | 16,334 | 19,225 | 0 |
| Jun | 0 | 2,764 | 5,407 | 10,037 | 13,253 | 16,054 | 18,996 | 0 |
| Jul | 0 | 3,064 | 5,685 | 9,906 | 13,960 | 17,298 | 20,010 | 0 |
| Aug | 371 | 3,072 | 5,935 | 10,381 | 13,883 | 16,666 | 18,458 | 0 |
| Sep | 1,411 | 3,214 | 5,945 | 10,193 | 13,431 | 16,974 | 19,439 | 0 |
| Oct | 1,423 | 3,429 | 5,994 | 9,842 | 14,645 | 18,339 | 19,623 | 0 |
| Nov | 1,265 | 3,733 | 6,801 | 10,993 | 14,990 | 18,187 | 19,601 | 0 |
| Dec | 1,159 | 3,538 | 6,588 | 10,110 | 13,279 | 16,528 | 17,441 | 0 |

**Table 5.2:** Stack Overflow Posts Per Day Per Month

## 2.2. DBA.STACKEXCHANGE

Even not busy as Stack Overflow as seen as in Table 5.3 and Table 5.4 Database Administrators community of Stack Exchange has more than 2,500 posts made by users per month which equals more than 80 questions and answers every day.

| DBA.StackExchange Posts Per Month | | | | | | | | |
|---|---|---|---|---|---|---|---|---|
| | 2008 | 2009 | 2010 | 2011 | 2012 | 2013 | 2014 | 2015 |
| Jan | | 1 | 1 | 9 | 725 | 1,624 | 1,589 | 2,708 |
| Feb | | 1 | 4 | 7 | 867 | 1,485 | 1,575 | 2,906 |
| Mar | | 4 | 7 | 14 | 856 | 1,693 | 1,789 | 2,481 |
| Apr | | 1 | 27 | 13 | 907 | 1,672 | 1,565 | 2,529 |
| May | | 8 | 13 | 28 | 1,099 | 1,564 | 1,602 | |
| Jun | | 3 | 4 | 713 | 1,433 | 1,557 | 2,129 | |
| Jul | 0 | 0 | 5 | 398 | 1,513 | 1,655 | 1,910 | |
| Aug | 0 | 0 | 9 | 403 | 1,555 | 2,137 | 2,102 | |
| Sep | 10 | 10 | 7 | 298 | 1,239 | 2,206 | 1,989 | |
| Oct | 10 | 7 | 14 | 513 | 1,283 | 2,172 | 2,173 | |
| Nov | 5 | 5 | 13 | 421 | 1,465 | 1,664 | 2,170 | |
| Dec | 0 | 5 | 28 | 477 | 1,491 | 1,736 | 2,515 | |

**Table 5.3:** Database Administrators Posts Per Month

| DBA.StackExchange Posts Per Day (Monthly) | | | | | | | | |
|---|---|---|---|---|---|---|---|---|
| | 2008 | 2009 | 2010 | 2011 | 2012 | 2013 | 2014 | 2015 |
| Jan | 0 | 0 | 0 | 0 | 23 | 52 | 51 | 87 |
| Feb | 0 | 0 | 0 | 0 | 30 | 53 | 56 | 104 |
| Mar | 0 | 0 | 0 | 0 | 28 | 55 | 58 | 80 |
| Apr | 0 | 0 | 1 | 0 | 30 | 56 | 52 | 84 |
| May | 0 | 0 | 0 | 1 | 35 | 50 | 52 | 0 |
| Jun | 0 | 0 | 0 | 24 | 48 | 52 | 71 | 0 |
| Jul | 0 | 0 | 0 | 13 | 49 | 53 | 62 | 0 |
| Aug | 0 | 0 | 0 | 13 | 50 | 69 | 68 | 0 |
| Sep | 0 | 0 | 0 | 10 | 41 | 74 | 66 | 0 |
| Oct | 0 | 0 | 0 | 17 | 41 | 70 | 70 | 0 |
| Nov | 0 | 0 | 0 | 14 | 49 | 55 | 72 | 0 |
| Dec | 0 | 0 | 1 | 15 | 48 | 56 | 81 | 0 |

**Table 5.4:** Database Administrators Posts Per Day Per Month

## 2.3. GITHUB

As seen as in Table 5.5 and Table 5.6 StackOverflow has more than 12,000,000 events arises every month which equals more than 400,000 events every day.

| GitHub Events Per Month | | | | |
|---|---|---|---|---|
| | 2012 | 2013 | 2014 | 2015 |
| Jan | | 5,209,199 | 8,903,382 | |
| Feb | | 5,385,789 | 9,113,031 | |
| Mar | 7,740,402 | 6,220,480 | 10,341,526 | |
| Apr | 10,589,082 | 6,391,854 | 10,663,768 | |
| May | 10,703,460 | 6,391,920 | 10,884,086 | |
| Jun | 8,876,202 | 6,228,482 | 10,135,128 | |
| Jul | 7,864,820 | 6,764,751 | 11,044,260 | |
| Aug | 11,050,352 | 6,926,377 | 11,056,705 | |
| Sep | 5,507,385 | 6,661,477 | 11,886,527 | |
| Oct | 4,434,096 | 8,322,995 | 13,121,784 | |
| Nov | 4,616,473 | 8,371,916 | 13,060,805 | |
| Dec | 4,331,680 | 7,817,420 | 12,424,903 | |

**Table 5.5:** GitHub Events Per Month

| GitHub Events Per Day (Monthly) | | | | |
|---|---|---|---|---|
| | 2012 | 2013 | 2014 | 2015 |
| Jan | 0 | 168,039 | 287,206 | 0 |
| Feb | 0 | 192,350 | 325,465 | 0 |
| Mar | 249,690 | 200,661 | 333,598 | 0 |
| Apr | 352,969 | 213,062 | 355,459 | 0 |
| May | 345,273 | 206,191 | 351,100 | 0 |
| Jun | 295,873 | 207,616 | 337,838 | 0 |
| Jul | 253,704 | 218,218 | 356,266 | 0 |
| Aug | 356,463 | 223,432 | 356,668 | 0 |
| Sep | 183,580 | 222,049 | 396,218 | 0 |
| Oct | 143,035 | 268,484 | 423,283 | 0 |
| Nov | 153,882 | 279,064 | 435,360 | 0 |
| Dec | 139,732 | 252,175 | 400,803 | 0 |

**Table 5.6:** GitHub Events Per Day Per Month

# 3. PROGRAMMING LANGUAGES

As everything became programmable Programming Language has become a tool or weapon of choice depends on the development will be made. Also not just context of the development effects the selected language. Also general trends of the software sector has an effect on the programming language selection. For example COBOL was the Data processing language for decades as an imperative and procedural language, but today most of the data processing application is developed using Functional or Object Oriented programing languages like R, Java or Python. In this context we selected 20 programming languages that widely in use today to analysis of their trends.

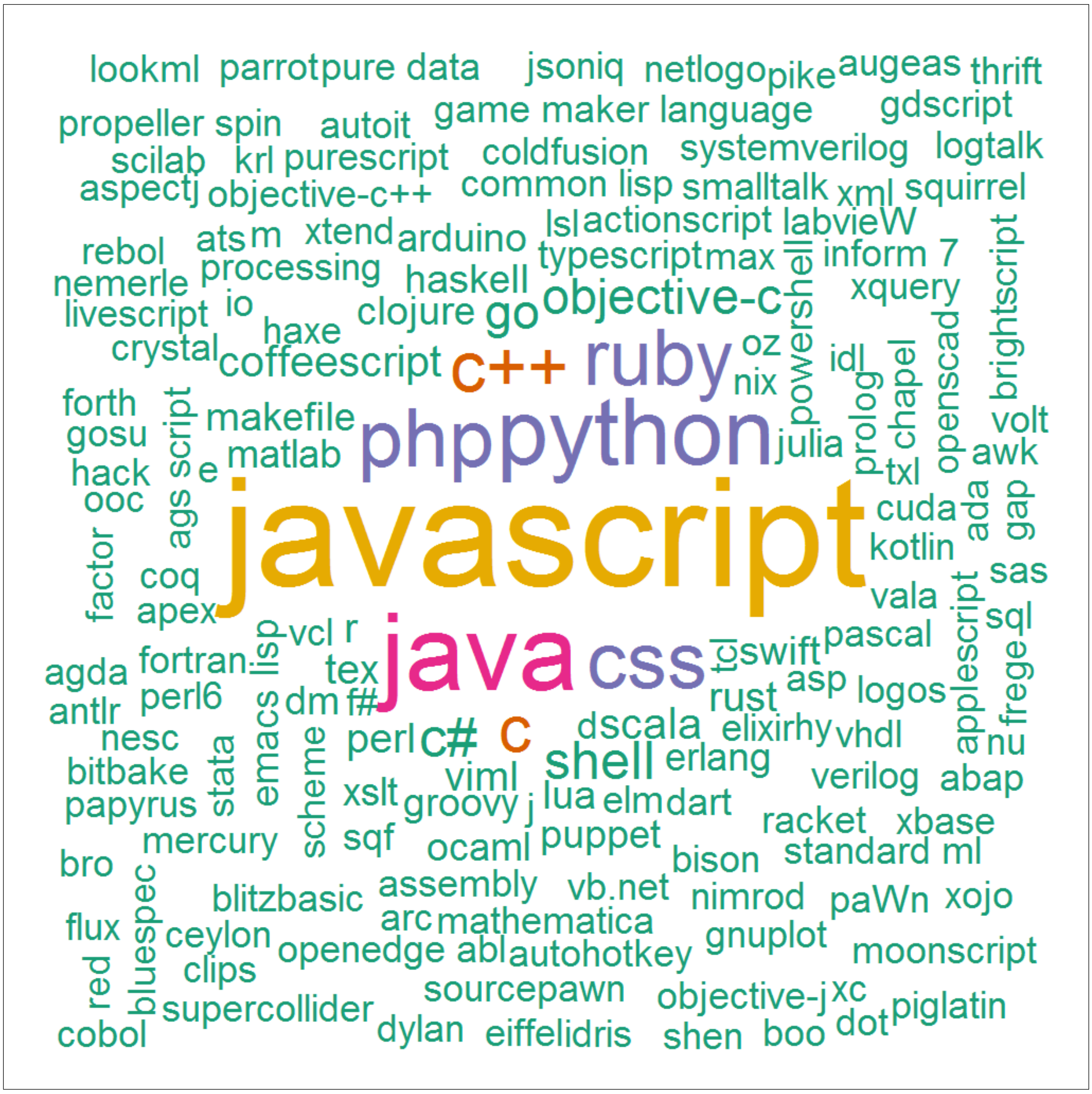

**Figure 5.1:** GitHub Events Per Day Per Month

Figure 5.1 could give some insight about the weight of the programming languages in computing world.

| Programming Languages | Keywords |
|---|---|
| JavaScript | JavaScript, js, ECMAScript, .js, JavaScript-execution, classic JavaScript, JavaScript-alert, JavaScript-DOM, JavaScript-disabled, JavaScript-library |
| Java | java-se, java, j2se, core-java, jdk, jre, java-libraries |
| Php | php, php-oop, php-date, php-frameworks, hypertext-preprocessor, php.ini, php-cli, php-errors, php-mail, php-cgi, php-functions, php-readfile, php-session |
| C# | c#, c-sharp, c#.net, c#-language, visual-c#, csharp |
| Python | python, pythonic, python-interpreter, python-shell |
| C++ | c++, cpp |
| SQL | sql, sql-query, sql-statement, sql-syntax, sql-select, sqlselect, select-statement |
| Swift | swift, swift-language, swift-ios, swift1.2 |
| R | r, rstats, r-language |
| Objective C | objective-c, objc |
| VB.NET | vb.net, vbproj, vb |
| Matlab | matlab, matlab-ide, matlab-path, matlab-toolbox |
| Scala | scala |
| PERL | perl |
| Haskell | haskell |
| Go | go, golang, go-language |
| Assembly | assembly, asm, assembler, assembly-language |
| XSLT | xslt, xsl, xsltransform, xsltprocessor, xsltproc, xslt-2.0-processors, xml-transform |
| Ruby | ruby, ruby-on-rails, ruby-on-rails-3, ruby-on-rails-4, ruby-on-rails-3.2, ruby-on-rails-3.1, rubygems, jruby, rubymine, ruby-on-rails-plugins, ruby-on-rails-2, rubymotion, ruby-1.9.3, ruby-on-rails-4.1, ruby-2.0, ruby-1.9, jrubyonrails, ironruby, ruby-on-rails-4.2, ruby-1.9.2, ruby-datamapper, macruby, ruby-1.8.7, ruby-2.1, mechanize-ruby, ruby-debug, ruby-1.8, ruby-c-extension, sqlite3-ruby, google-api-ruby-client, rubyzip, therubyracer, bcrypt-ruby, ruby-enterprise-edition, qtruby, mongodb-ruby |

**Table 5.7:** Programming Languages and Their Keywords

# 4. DATABASES

Databases are very wide and improving area of Information Sector as Data become the center of the modern life. RDBM Systems were the biggest change in the database technology until decade. Today we many Database solutions to different development contexts. NoSQL database technologies like Column Stores, Document Stores, Key Value Stores, and Graph Stores has increase their user base every day. In this context we have selected 8 different database server to analyze their trends.

| Databases | Keywords |
|---|---|
| Microsoft SQL Server | sql-server, sql-server-2008, sql-server-2005, sql-server-2008-r2, sql-server-2012, sql-server-ce, sql-server-2000, sql-server-express, sql-server-2014, sql-server-2012-express, sql-server-2008-express, sql-server-triggers, sql-server-ce-4, sql-server-agent, sql-server-2008r2-express, sql-server-2005-express, sql-server-express-2008, sql-server-2014-express, sql-server-profiler, sql-server-ce-3.5, sql-server-2012-datatools, sql-server-group-concat, sql-server-native-client, sql-server-data-tools |
| MySQL | mysql, mysqli, mysql-workbench, mysqldump, mysql-python, mysql-error-1064, mysql-connector, mysql2, mysql-real-escape-string, node-mysql, mysql-management, mysql-error-1054, mysql5, mysql-num-rows, mysql-connect, rmysql, mysql-error-1045, mysql-error-1005, mysql-5.6, mysqlimport, mysql-5.5, pymysql, mysql-error-1062, mysql-insert-id, mysql-cluster, mysql-5.1, mysqli-multi-query, mysql-slow-query-log, mysql-error-1093, mysql++, mysql-error-1111, libmysql, mysqlconnection, mysql-error-1242, mysqlnd, mysql-error-1146 |
| Oracle | oracle, oracle11g, oracle10g, oracle-sqldeveloper, oracle-apex, oracle-adf, oracle9i, oracleforms, cx-oracle, oracle12c, oracle11gr2, oracle-xe, oracle-coherence, oracleclient, oraclereports, oracle-pro-c, oracle-spatial, oracle-aq, oracle-apps, oracle8i, oracle-adf-mobile, oracle-text, oracle-fusion-middleware, oracle-ucm, oracle-service-bus, oracle-data-integrator, oracle-maf, oracleapplications, oracle-soa, system.data.oracleclient, oraclecommand, oracle-commerce, oracle-golden-gate, oracle-sql-data-modeler, oracle-application-server, oracle-xml-db |
| MongoDB | mongodb, mongodb-query, mongodb-csharp, spring-data-mongodb, mongodb-java, mongodb-php, node-mongodb-native, mongodb-indexes, rmongodb, django-mongodb-engine, doctrine-mongodb, mongodb-ruby, mongodb-mms, mongodb-update, mongodb-c, mongodb-shell, mongodb-scala, gorm-mongodb, objcmongodb, jenssegers-mongodb, mongodb-csharp-2.0, mongodb-hadoop, spring-mongodb, mongodb-roles, mongodb.driver, mongodb-rest |
| DB2 | db2, db2400, db2-luw, mdb2, db2-connect, db2i, ibm-db2, db2-content-manager |
| PostgreSQL | postgresql, postgresql-9.1, postgresql-9.2, postgresql-9.3, rails-postgresql, postgresql-8.4, postgresql-performance, postgresql-9.4, postgresql-9.0, django-postgresql, rpostgresql, postgresql-8.3, postgresql-8.2, sails-postgresql, postgresql-json, postgresql-copy, postgresql-8.1, ef-postgresql, py-postgresql, postgresql-initdb, postgresql-extensions, hibernate-postgresql, postgresql-8.0 |
| Microsoft Access | ms-access, ms-access-2007, ms-access-2010, ms-access-2013, ms-access-2003, ms-access-2000, ms-access-97, ms-access-data-macro, ms-access-2002 |
| Amazon RDS | amazon-rds |

**Table 5.8:** Databases and Their Keywords

# 5. CLOUD SERVICES

Cloud computing is one of the important trend in information technology. Every technology company uses fruits of the cloud computing in their products. Even they don’t provide any cloud service like Amazon or Microsoft every big company have some product that works in the cloud. In this context we have selected 5 different cloud service to analyze their trends.

| Cloud Services | Keywords |
|---|---|
| Amazon Web Services | amazon, amazon-web-services, amazon-ec2, amazon-s3, amazon-dynamodb, amazon-cloudfront, amazon-rds, amazon-redshift, amazon-sqs, amazon-simpledb, amazon-ses, amazon-emr, amazon-sns, amazon-elb, amazon-route53, amazon-cloudformation, amazon-mws, amazon-ebs, amazon-vpc, amazon-product-api, amazon-cloudwatch, amazon-iam, amazon-elasticache, amazon-cognito, amazon-cloudsearch, amazon-swf, amazon-appstore, amazon-data-pipeline, amazon-glacier, amazon-kinesis, amazon-elastic-beanstalk, knox-amazon-s3-client, amazon-fire-tv, amazon-ecs, amazon-fps, amazon-payments |
| Google Cloud | google-cloud-storage, google-cloud-endpoints, google-cloud-sql, google-cloud-messaging, google-cloud-platform, google-cloud-datastore, google-cloud-dataflow, google-cloud-print, google-cloud-dns, google-cloud-console, google-cloud-pubsub, google-cloud-bigtable, google-cloud-debugger, google-cloud-save, google-cloud-trace, google-app-engine, google-bigquery, google-compute-engine |
| Microsoft Azure | azure, sql-azure, windows-azure-storage, azure-mobile-services, azure-web-sites, azure-storage-blobs, azure-web-roles, azure-worker-roles, azure-storage-tables, azureservicebus, azure-active-directory, azure-virtual-machine, azure-webjobs, azure-table-storage, azure-cloud-services, azure-caching, azure-media-services, azure-documentdb, azure-powershell, azure-diagnostics, azure-notificationhub, azure-sdk-.net, azure-servicebus-queues, windows-azure-queues, azure-deployment, windows-azure-diagnostics, azure-webjobssdk, azure-configuration, azure-compute-emulator, azure-virtual-network, azure-search, azure-vm-role, azure-appfabric, azure-cdn, azure-acs, azure-servicebusrelay |
| Heroku | heroku, heroku-toolbelt, heroku-postgres, heroku-san, heroku-api |

**Table 5.9:** Cloud Services and Their Keywords

# 6. MOBILE OPERATING SYSTEMS

As technology advances ability of reaching information on the move become ordinary. Greatest share of this improvements is belong to the smart phones. As those smart phones getting powerful everyday their operating systems getting more complicated. Belong all those mobile operating systems we have selected Android, IOS and Windows Phone for our research.

| Mobile Operating Systems | Keywords |
|---|---|
| Android | android, android-layout, android-intent, android-fragments, android-activity, android-listview, android-studio, android-asynctask, android-ndk, android-emulator, android-actionbar, android-edittext, android-viewpager, android-widget, android-webview, android-service, android-camera, android-gcm×, android-sqlite, android-manifest, android-linearlayout, android-animation, android-canvas, android-arrayadapter, android-imageview, android-mediaplayer, google-maps-android-api-2, android-gradle, android-view, android-textview, android-contentprovider, android-tabhost, android-5.0-lollipop, android-mapview, android-spinner, android-notifications |
| IOS | ios, ios7, ios5, ios4, ios6, ios8, ios-simulator, facebook-ios-sdk, google-maps-sdk-ios, ios7.1, ios-ui-automation, ios-provisioning, ios8.1, ios-app-extension, ios-4.2, ios5.1, ios-universal-app, ios6.1, ios8-extension, ios-frameworks, ios8-today-widget, ios8.3, ios6-maps, ios-autolayout |
| Windows Phone | windows-phone, windows-phone-7, windows-phone-8, windows-phone-8.1, windows-phone-7.1, windows-phone-8-emulator, windows-phone-7.1.1, windows-phone-7-emulator, windows-phone-8-sdk, windows-phone-toolkit, windows-phone-store, windows-phone-7.8, windows-phone-voip, windows-phone-emulator, windows-phone-sl-8.1 |

**Table 5.10:** Mobile Operating Systems and Their Keywords

This Page Intentionally Left Blank

# 6 EXPERIMENTAL RESULTS

We have analyzed our data using relation graphs and time series forecasts. Data of relation graphs are created yearly using discretization. Using this discrete data we have visualized this data in Gephi for years between 2009 and 2014. Exploratory analysis made to those visualized data but analysis of metrics left for a future research.

Time series of the technologies are forecasted using “forecast” library of the R. Among the methods that present in this library we have selected Exponential Smoothing (ETS) method to calculate predictions.

Data is used to create 3 different forecasts for each of Programming Language Knowledge Area and 2 forecasts for other three knowledge areas. StackExchange data is used for creating quarterly and monthly forecasts. While creating both forecasts based on StackExchange data we have used data from 2009 to 2014 as training data and data from the January of 2014 to February of 2015 as test data.

Forecasts relied on GitHub data is created only for Programming Languages Knowledge Area. Because of small size of GitHub dataset (it is starting from 2012) we cannot create meaningful forecast in quarterly basis. As we want to increase success of our forecast we have applied monthly forecasts in GitHub data. For example for forecast of November 2014 we have used data from January 2012 to October 2014 included.

# 1. PROGRAMMING LANGUAGES

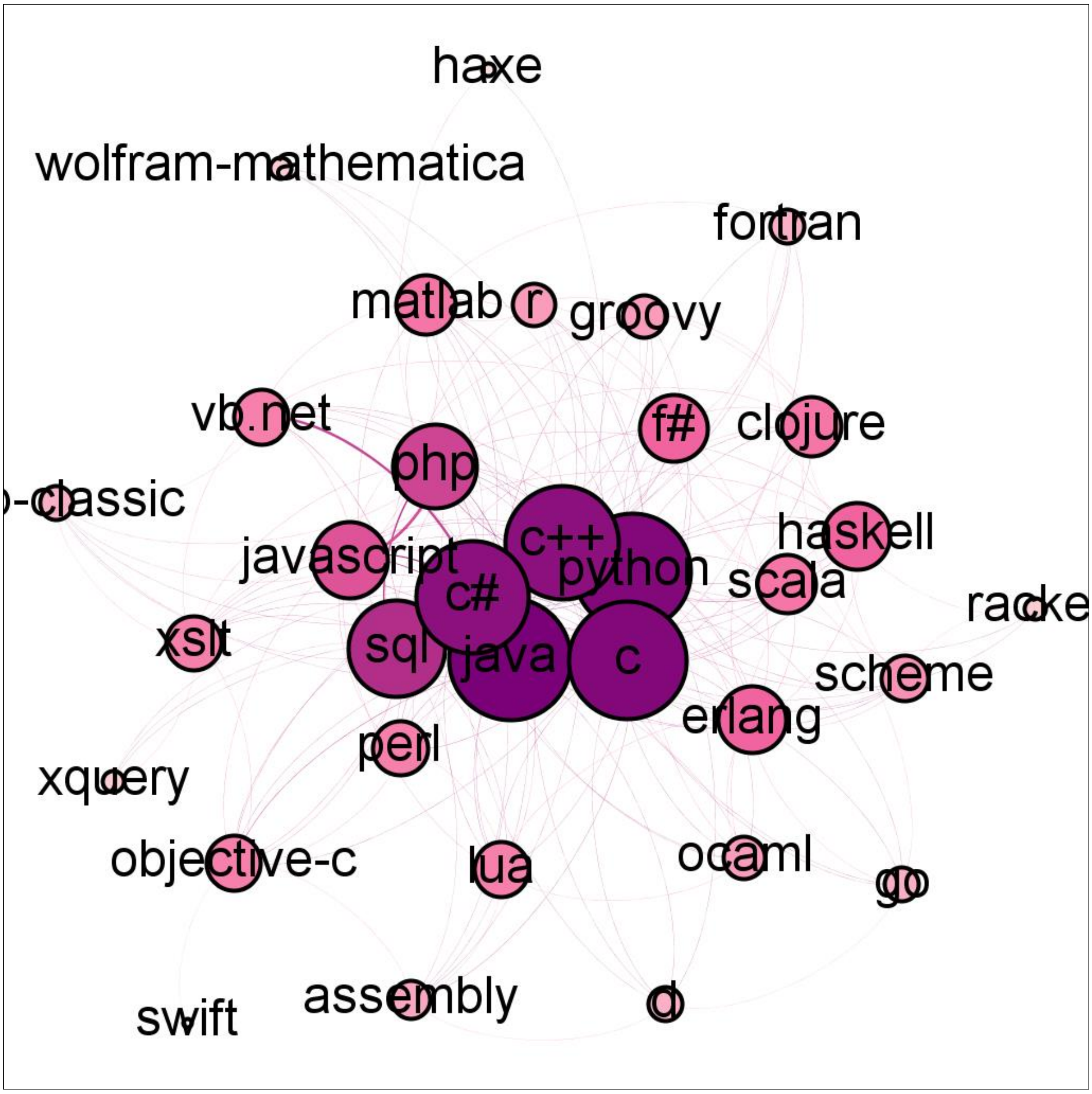

***Figure 6.1:*** *Relations of Programing Languages – 2009*

Exploratory analysis of the graph shows us 5 mainstream languages are C, C++, C#, Java and Python. PHP, SQL and JavaScript are secondary rank languages. Clojure, Erlang, Scala and Haskell are prominent functional languages. Another notable language is R which is even smaller weight than Matlab in 2009.

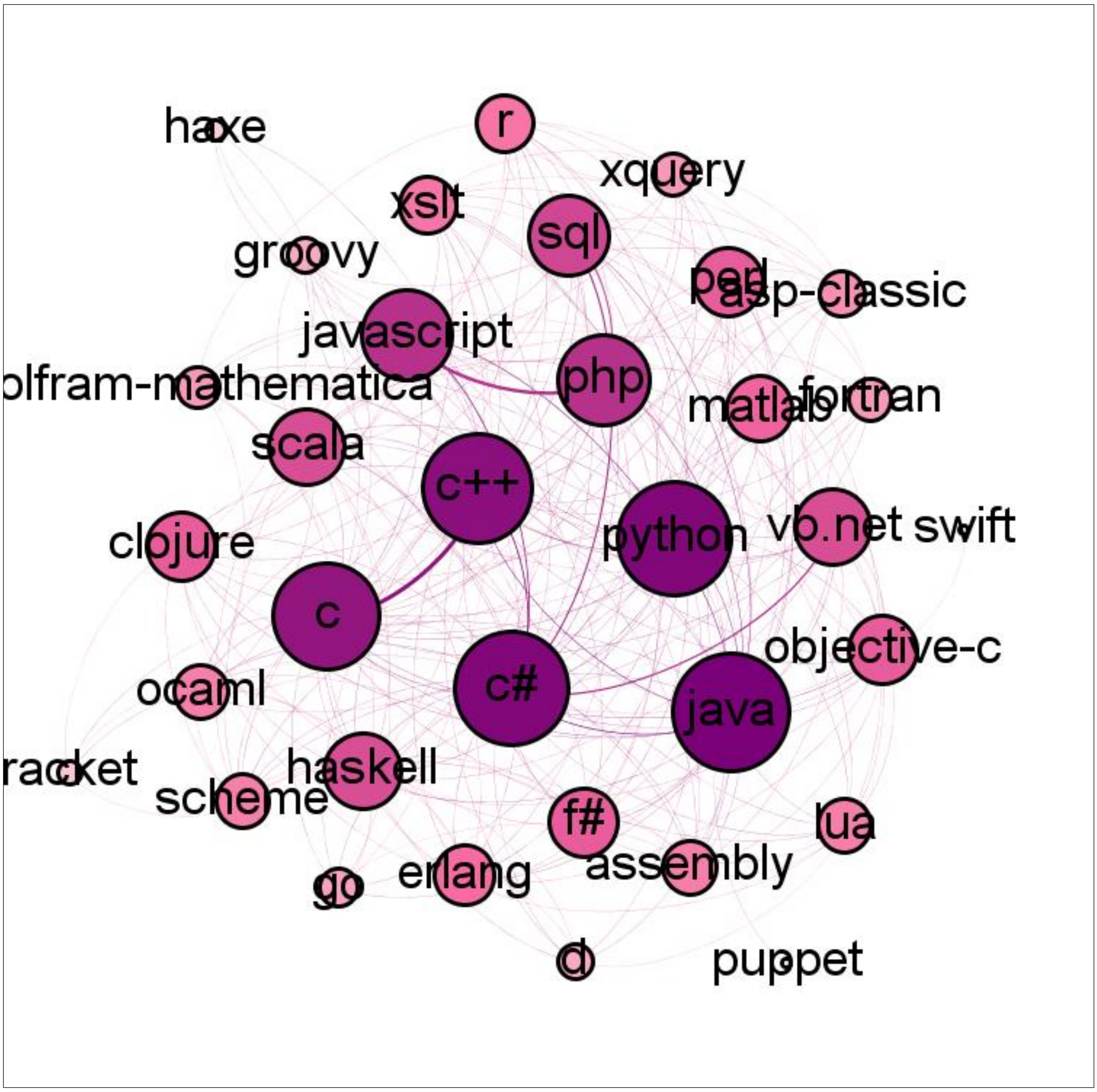

***Figure 6.2:*** *Relations of Programing Languages – 2010*

Exploratory analysis of the graph shows us C, C++, C#, Java and Python are still mainstream languages. JavaScript and PHP increases their weight. Weight of the SQL is lowering. Scala and Haskell rising between functional languages. Both Matlab and R increases their weight as numerical analysis and data analysis languages. Another notable difference is rise of Objective C as increased earnings in App Store [27].

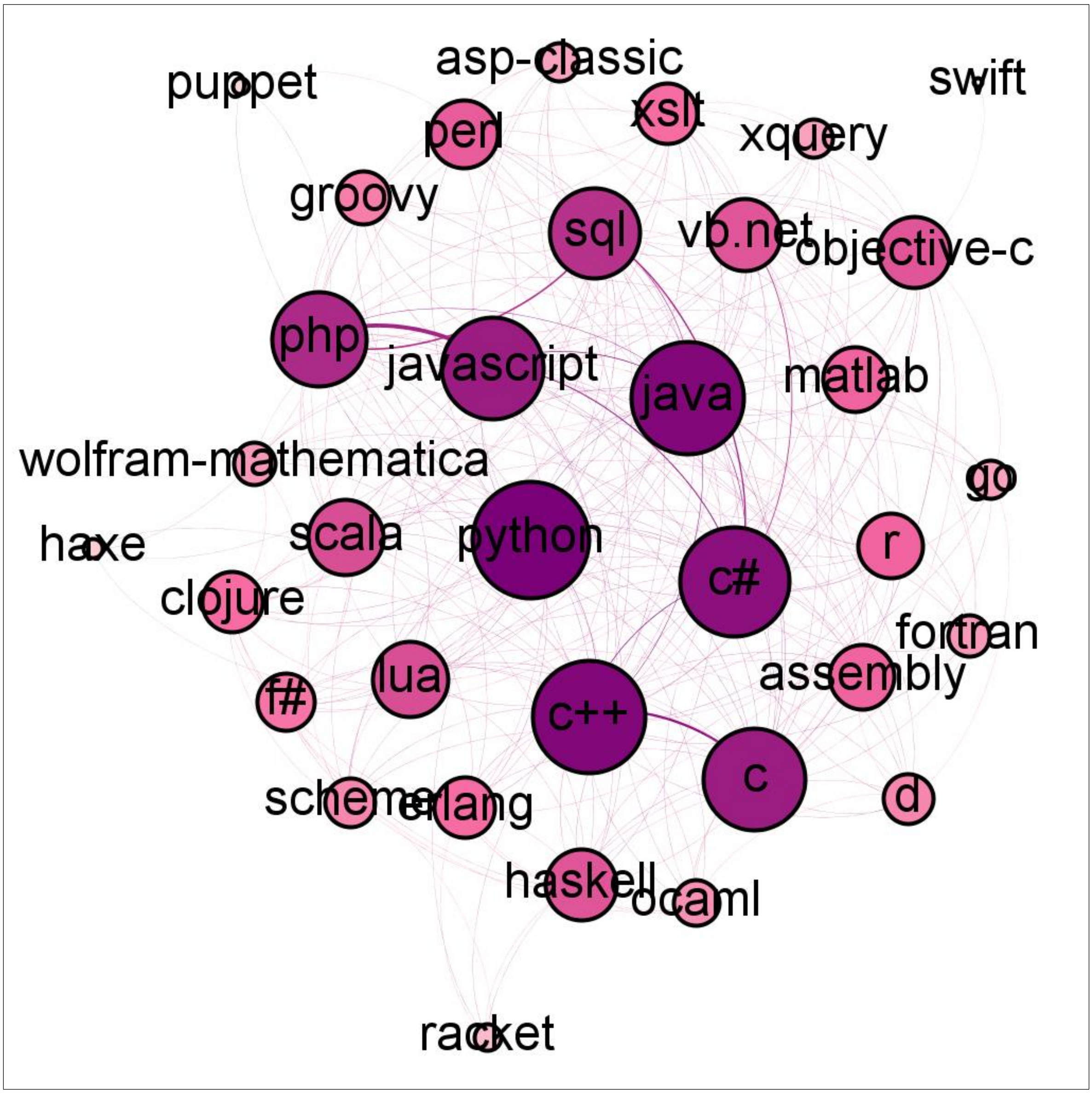

***Figure 6.3:*** *Relations of Programing Languages – 2011*

Exploratory analysis of the graph shows us C, C++, C#, Java and Python joined by PHP and JavaScript in mainstream level. Also C loses its relative weight even it is still main stream. Relative weight of the SQL is increasing. Scala and Haskell still top of the functional languages. While R increases its relative weight Matlab and R still top as numerical analysis and data analysis languages.

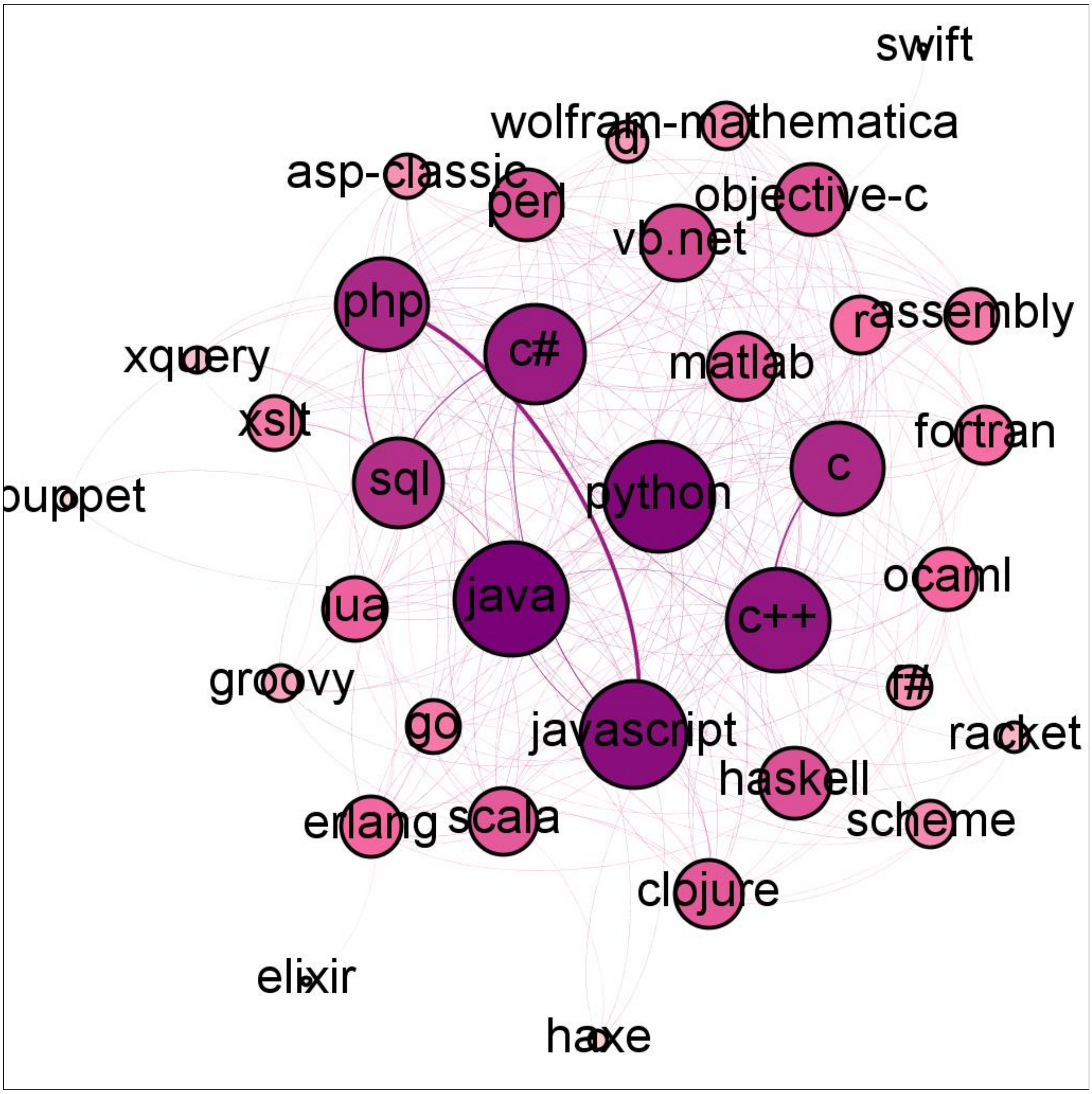

***Figure 6.4:*** *Relations of Programing Languages – 2012*

Exploratory analysis of the graph shows us C, C++, C#, Java, Python, PHP and JavaScript in mainstream level. Also still C loses its relative weight even it is still main stream. Relative weight of the SQL is lowering. Clojure and Erlang catches Scala and Haskell at top of the functional languages. While Matlab increases its relative weight Matlab and R still top as numerical analysis and data analysis languages.

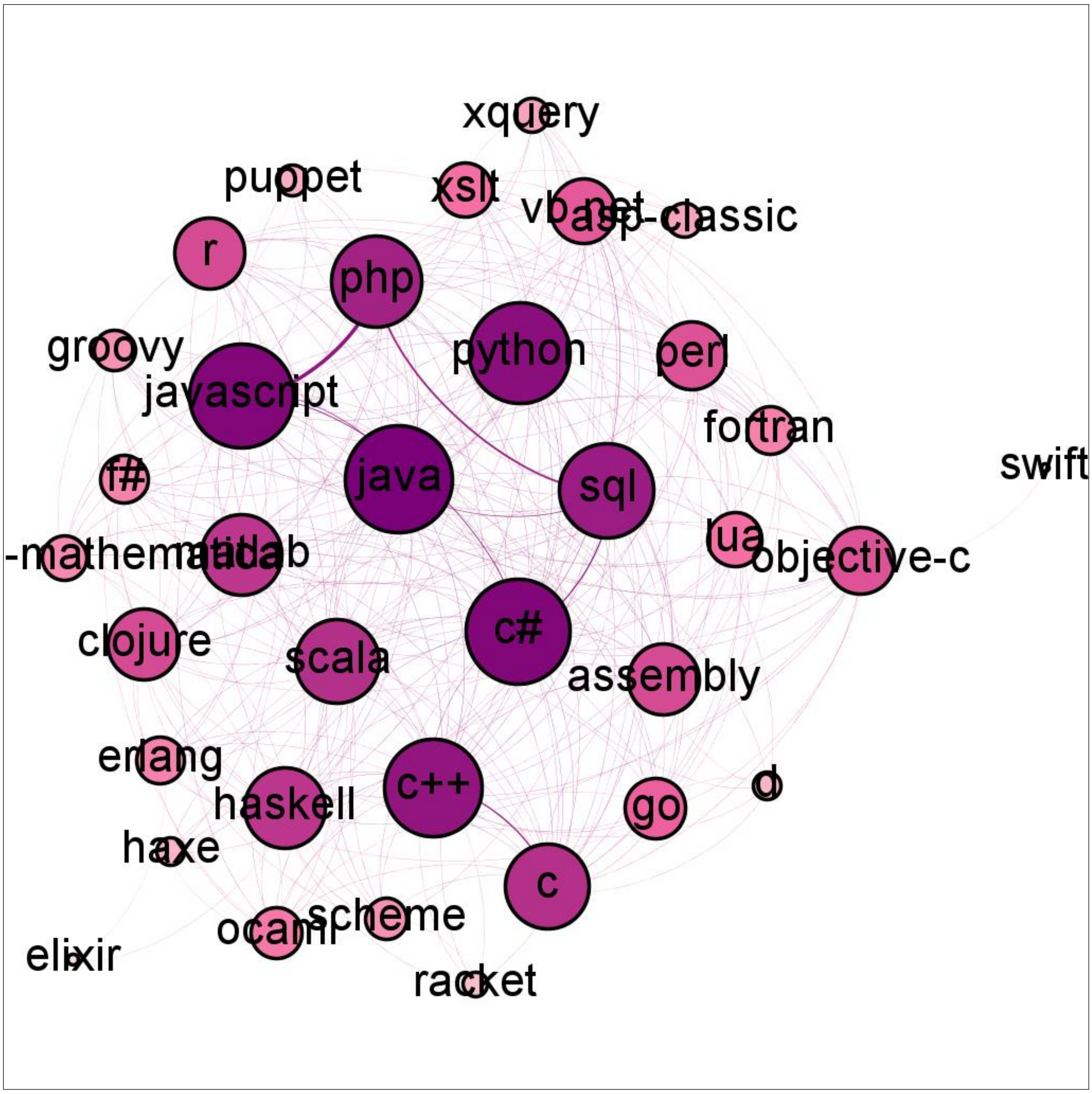

***Figure 6.5:*** *Relations of Programing Languages – 2013*

Exploratory analysis of the graph shows us C++, C#, Java, Python, PHP, SQL and JavaScript in mainstream level. As C loses its relative weight and SQL increases, SQL gets place of C in main stream languages. Scala, Haskell, Matlab and R increases their relative weight. This increase in functional, numerical analysis and data analysis languages maybe lead a paradigm shift in programming languages.

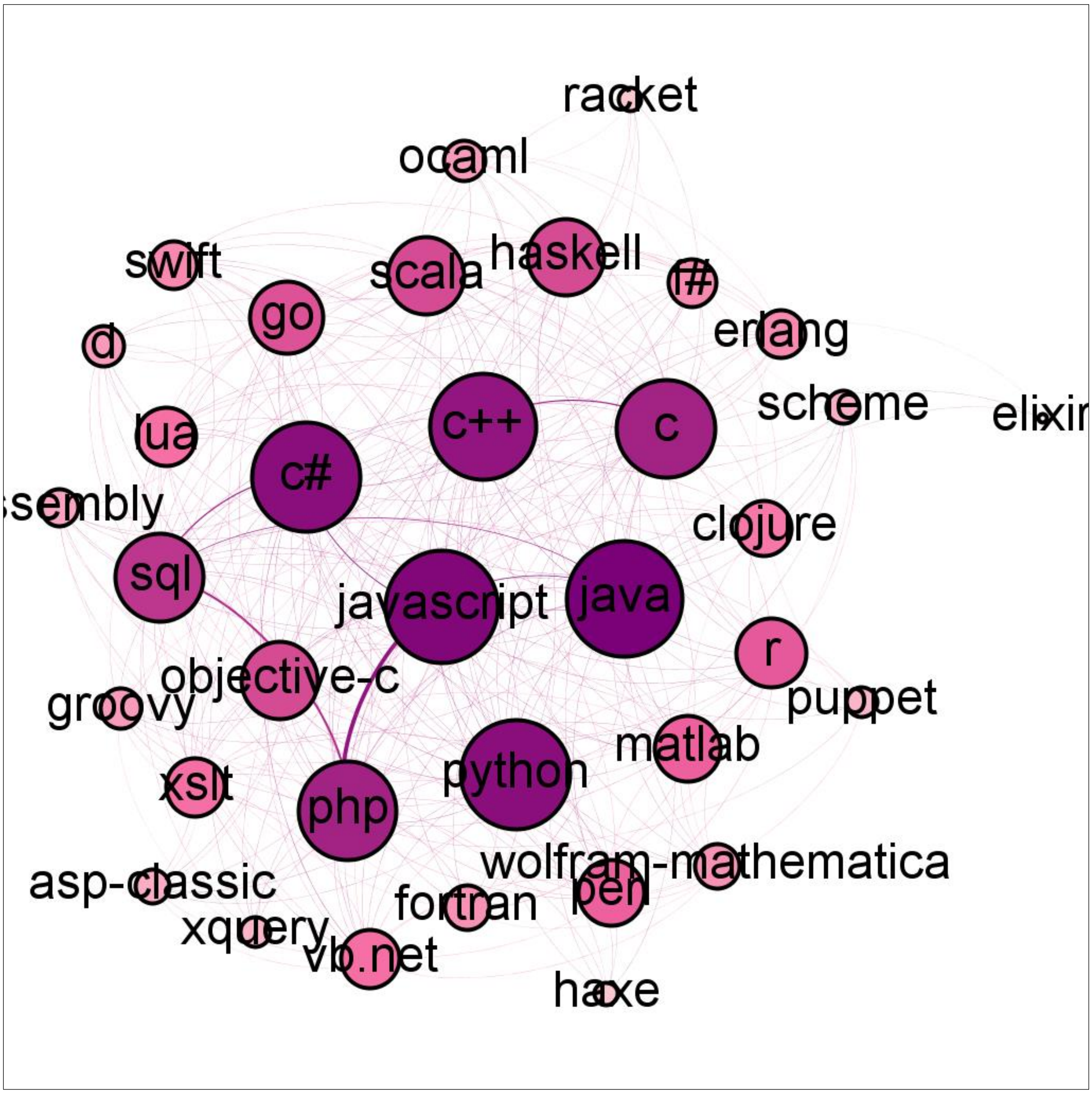

***Figure 6.6:*** *Relations of Programing Languages – 2014*

Exploratory analysis of the graph shows us C returns back to main stream level and joined C++, C#, Java, Python, PHP, SQL and JavaScript. Scala and Haskell stays top of the functional languages. Matlab and R is stays top of numerical analysis and data analysis languages. Also huge increase in relative weights of Swift and Objective C enormous.

## 1.1. JAVASCRIPT

Tables 6.1, 6.2 and 6.3 shows all three forecast created shows results a little less than actual results. This might be a result huge rise of JavaScript with micro services and node.js. But still not much to talk about as results and predictions are not so different than each other.

| StackOverflow JavaScript Actual-Prediction (Quarterly) (% Error) | | | |
|---|---|---|---|
| | Actual | Prediction | % Error |
| 2014 Q1 | 67254 | 60927.84 | 9.41% |
| 2014 Q2 | 66484 | 65320.64 | 1.75% |
| 2014 Q3 | 63406 | 69713.45 | 9.95% |
| 2014 Q4 | 60688 | 74106.26 | 22.11% |

***Table 6.1:*** *Quarterly Relative error of JavaScript Prediction on StackExchange Data*

$$MRE = \frac{1}{N}\sum_{1}^{N}\frac{A_i - P_i}{A_i} = \%10.80$$

$$MdRE = \%9.68$$

| StackOverflow JavaScript Actual-Prediction (Monthly) (% Error) | | | |
|---|---|---|---|
| | Actual | Prediction | % Error |
| Jan-14 | 21437 | 19134.96 | 10.74% |
| Feb-14 | 21492 | 19468.58 | 9.41% |
| Mar-14 | 24325 | 19796.97 | 18.61% |
| Apr-14 | 23879 | 20120.05 | 15.74% |
| May-14 | 22097 | 20437.72 | 7.51% |
| Jun-14 | 20508 | 20749.93 | 1.18% |
| Jul-14 | 22385 | 21056.60 | 5.93% |
| Aug-14 | 20728 | 21357.71 | 3.04% |
| Sep-14 | 20293 | 21653.19 | 6.70% |
| Oct-14 | 21391 | 21943.04 | 2.58% |
| Nov-14 | 20366 | 22227.24 | 9.14% |
| Dec-14 | 18931 | 22505.77 | 18.88% |
| Jan-15 | 20780 | 22778.65 | 9.62% |
| Feb-15 | 21568 | 23045.87 | 6.85% |

***Table 6.2:*** *Monthly Relative error of JavaScript Prediction on StackExchange Data*

$$MRE = \frac{1}{N}\sum_{1}^{N}\frac{A_i - P_i}{A_i} = \%9.00$$

$$MdRE = \%7.18$$

| GitHub JavaScript Actual-Prediction (Monthly) (% Error) | | | |
|---|---|---|---|
| | Actual | Prediction | % Error |
| Jan-14 | 1562149 | 1384391.00 | 11.38% |
| Feb-14 | 1543424 | 1550369.00 | 0.45% |
| Mar-14 | 1784465 | 1543885.00 | 13.48% |
| Apr-14 | 1839167 | 1768510.00 | 3.84% |
| May-14 | 1897223 | 1835182.00 | 3.27% |
| Jun-14 | 1810145 | 1893842.00 | 4.62% |
| Jul-14 | 1987131 | 1814990.00 | 8.66% |
| Aug-14 | 2013961 | 1976309.00 | 1.87% |
| Sep-14 | 2045347 | 2011668.00 | 1.65% |
| Oct-14 | 2242171 | 2126321.00 | 5.17% |
| Nov-14 | 2249222 | 2320122.00 | 3.15% |
| Dec-14 | 2160044 | 2404472.00 | 11.32% |

***Table 6.3:*** *Monthly Relative error of JavaScript Prediction on GitHub Data*

$$MRE = \frac{1}{N}\sum_{1}^{N}\frac{A_i - P_i}{A_i} = \%5.74$$

$$MdRE = \%4.23$$

## 1.2. JAVA

Tables 6.4 and 6.5 are shows Java mentioned less than expected in programing communities according to our forecast. Also as seen in Table 6.6 events related to Java is less than expected in GitHub. But still not much to talk about as results and predictions are not so different than each other.

| StackOverflow Java Actual-Prediction (Quarterly) (% Error) | | | |
|---|---|---|---|
| | Actual | Prediction | % Error |
| 2014 Q1 | 62364 | 61455.79 | 1.46% |
| 2014 Q2 | 61622 | 63420.09 | 2.92% |
| 2014 Q3 | 55948 | 65126.49 | 16.41% |
| 2014 Q4 | 58636 | 67816.51 | 15.66% |

***Table 6.4:*** *Quarterly Relative error of Java Prediction on StackExchange Data*

$$MRE = \frac{1}{N}\sum_{1}^{N}\frac{A_i - P_i}{A_i} = \%9.11$$

$$MdRE = \%9.29$$

| StackOverflow Java Actual-Prediction (Monthly) (% Error) | | | |
|---|---|---|---|
| | Actual | Prediction | % Error |
| Jan-14 | 19024 | 21292.76 | 11.93% |
| Feb-14 | 19718 | 20674.59 | 4.85% |
| Mar-14 | 23622 | 23649.36 | 0.12% |
| Apr-14 | 23426 | 21922.43 | 6.42% |
| May-14 | 20210 | 22183.61 | 9.77% |
| Jun-14 | 17986 | 21094.15 | 17.28% |
| Jul-14 | 19514 | 21708.76 | 11.25% |
| Aug-14 | 17600 | 21290.82 | 20.97% |
| Sep-14 | 18834 | 20880.11 | 10.86% |
| Oct-14 | 20126 | 22622.28 | 12.40% |
| Nov-14 | 20172 | 23380.71 | 15.91% |
| Dec-14 | 18338 | 21868.33 | 19.25% |
| Jan-15 | 18488 | 26012.36 | 40.70% |
| Feb-15 | 19586 | 25174.06 | 28.53% |

**Table 6.5:** Monthly Relative error of Java Prediction on StackExchange Data

$$MRE = \frac{1}{N}\sum_{1}^{N}\frac{A_i - P_i}{A_i} = \%15.02$$

$$MdRE = \%14.16$$

| GitHub Java Actual-Prediction (Monthly) (% Error) | | | |
|---|---|---|---|
| | Actual | Prediction | % Error |
| Jan-14 | 891887 | 808414.80 | 9.36% |
| Feb-14 | 910866 | 891878.60 | 2.08% |
| Mar-14 | 1080747 | 910864.10 | 15.72% |
| Apr-14 | 1107541 | 1080730.00 | 2.42% |
| May-14 | 1181326 | 1107538.00 | 6.25% |
| Jun-14 | 1045039 | 1181319.00 | 13.04% |
| Jul-14 | 1065219 | 1045053.00 | 1.89% |
| Aug-14 | 1056627 | 1065217.00 | 0.81% |
| Sep-14 | 1162541 | 1056628.00 | 9.11% |
| Oct-14 | 1333633 | 1162530.00 | 12.83% |
| Nov-14 | 1398060 | 1333616.00 | 4.61% |
| Dec-14 | 1296029 | 1398054.00 | 7.87% |

***Table 6.6:*** *Quarterly Relative error of JavaScript Prediction on StackExchange Data*

$$MRE = \frac{1}{N}\sum_{1}^{N}\frac{A_i - P_i}{A_i} = \%7.17$$

$$MdRE = \%7.06$$

## 1.3. PHP

Table 6.7 and 6.8 shows PHP is mentioned less than expected in programming communities according to our forecast. This might be a sign of a decrease of popularity of PHP. Because nearly actual count of posts that mentioned PHP is 25 percent less than expected in our forecast in 4th quarter of 2014. Table 6.9 show PHP usage is not significantly dropped in GitHub as opposite of Stack Overflow.

| StackOverflow PHP Actual-Prediction (Quarterly) (% Error) | | | |
|---|---|---|---|
| | Actual | Prediction | % Error |
| 2014 Q1 | 54373 | 48771.37 | 10.30% |
| 2014 Q2 | 51392 | 51133.87 | 0.50% |
| 2014 Q3 | 46604 | 53496.38 | 14.79% |
| 2014 Q4 | 44571 | 55858.88 | 25.33% |

***Table 6.7:*** *Quarterly Relative error of PHP Prediction on StackExchange Data*

$$MRE = \frac{1}{N}\sum_{1}^{N}\frac{A_i - P_i}{A_i} = \%12.73$$

$$MdRE = \%12.55$$

| StackOverflow PHP Actual-Prediction (Monthly) (% Error) | | | |
|---|---|---|---|
| | Actual | Prediction | % Error |
| Jan-14 | 17385 | 15399.18 | 11.42% |
| Feb-14 | 17103 | 15625.14 | 8.64% |
| Mar-14 | 19885 | 15851.10 | 20.29% |
| Apr-14 | 19871 | 16077.06 | 19.09% |
| May-14 | 16678 | 16303.02 | 2.25% |
| Jun-14 | 14843 | 16528.98 | 11.36% |
| Jul-14 | 16272 | 16754.94 | 2.97% |
| Aug-14 | 15006 | 16980.89 | 13.16% |
| Sep-14 | 15326 | 17206.85 | 12.27% |
| Oct-14 | 15577 | 17432.81 | 11.91% |
| Nov-14 | 15074 | 17658.77 | 17.15% |
| Dec-14 | 13920 | 17884.73 | 28.48% |
| Jan-15 | 15454 | 18110.69 | 17.19% |
| Feb-15 | 16005 | 18336.65 | 14.57% |

***Table 6.8:*** *Monthly Relative error of PHP Prediction on StackExchange Data*

$$MRE = \frac{1}{N}\sum_{1}^{N}\frac{A_i - P_i}{A_i} = \%13.63$$

$$MdRE = \%13.86$$

| GitHub PHP Actual-Prediction (Monthly) (% Error) | | | |
|---|---|---|---|
| | Actual | Prediction | % Error |
| Jan-14 | 696049 | 561564.70 | 19.32% |
| Feb-14 | 673980 | 696035.50 | 3.27% |
| Mar-14 | 743695 | 673982.20 | 9.37% |
| Apr-14 | 723922 | 743688.00 | 2.73% |
| May-14 | 734319 | 723924.00 | 1.42% |
| Jun-14 | 689784 | 734318.00 | 6.46% |
| Jul-14 | 742613 | 689788.50 | 7.11% |
| Aug-14 | 758334 | 742607.70 | 2.07% |
| Sep-14 | 776562 | 758332.40 | 2.35% |
| Oct-14 | 846312 | 776560.20 | 8.24% |
| Nov-14 | 850841 | 846305.00 | 0.53% |
| Dec-14 | 812482 | 850840.50 | 4.72% |

***Table 6.9:*** *Quarterly Relative error of PHP Prediction on StackExchange Data*

$$MRE = \frac{1}{N}\sum_{1}^{N}\frac{A_i - P_i}{A_i} = \%5.63$$

$$MdRE = \%4.0$$

## 1.4. C#

Table 6.10, 6.11 and 6.12 shows usage of C# is not much as expected in our forecasts. This might be the result of sharp rise of popularity of C# caused by technologies like MVC and Entity Framework in 2012 and 2013. We may tell more about the future of C# future years. Because with additions C# is very mature language as it is already in 6th release. Time will show us if C# will take more ground from Java in Enterprise world.

| StackOverflow C# Actual-Prediction (Quarterly) (% Error) | | | |
|---|---|---|---|
| | Actual | Prediction | % Error |
| 2014 Q1 | 50799 | 47878.18 | 5.75% |
| 2014 Q2 | 47820 | 48964.41 | 2.39% |
| 2014 Q3 | 46229 | 49952.40 | 8.05% |
| 2014 Q4 | 42400 | 50848.97 | 19.93% |

***Table 6.10:*** *Quarterly Relative error of C# Prediction on StackExchange Data*

$$MRE = \frac{1}{N}\sum_{1}^{N}\frac{A_i - P_i}{A_i} = \%9.03$$

$$MdRE = \%6.90$$

| StackOverflow C# Actual-Prediction (Monthly) (% Error) | | | |
|---|---|---|---|
| | Actual | Prediction | % Error |
| Jan-14 | 16725 | 16793.28 | 0.41% |
| Feb-14 | 16425 | 16768.73 | 2.09% |
| Mar-14 | 17649 | 19190.47 | 8.73% |
| Apr-14 | 16708 | 18213.67 | 9.01% |
| May-14 | 16181 | 18469.58 | 14.14% |
| Jun-14 | 14931 | 18468.11 | 23.69% |
| Jul-14 | 16390 | 19591.84 | 19.54% |
| Aug-14 | 14959 | 19383.87 | 29.58% |
| Sep-14 | 14880 | 17977.05 | 20.81% |
| Oct-14 | 15087 | 19005.26 | 25.97% |
| Nov-14 | 14225 | 18888.36 | 32.78% |
| Dec-14 | 13088 | 17498.92 | 33.70% |
| Jan-15 | 13968 | 19766.78 | 41.51% |
| Feb-15 | 14137 | 19694.72 | 39.31% |

***Table 6.11:*** *Monthly Relative error of C# Prediction on StackExchange Data*

$$MRE = \frac{1}{N}\sum_{1}^{N}\frac{A_i - P_i}{A_i} = \%21.52$$

$$MdRE = \%24.83$$

| GitHub C# Actual-Prediction (Monthly) (% Error) | | | |
|---|---|---|---|
| | Actual | Prediction | % Error |
| Jan-14 | 190330 | 167441.00 | 12.03% |
| Feb-14 | 189842 | 190327.70 | 0.26% |
| Mar-14 | 212447 | 189842.00 | 10.64% |
| Apr-14 | 232858 | 212444.70 | 8.77% |
| May-14 | 264093 | 232855.90 | 11.83% |
| Jun-14 | 235088 | 264089.90 | 12.34% |
| Jul-14 | 254236 | 235090.90 | 7.53% |
| Aug-14 | 261605 | 254234.10 | 2.82% |
| Sep-14 | 266484 | 261604.30 | 1.83% |
| Oct-14 | 317892 | 266483.50 | 16.17% |
| Nov-14 | 341784 | 331209.90 | 3.09% |
| Dec-14 | 319776 | 344531.70 | 7.74% |

**Table 6.12:** Quarterly Relative error of C# Prediction on StackExchange Data

$$MRE = \frac{1}{N}\sum_{1}^{N}\frac{A_i - P_i}{A_i} = \%7.92$$

$$MdRE = \%8.25$$

## 1.5. PYTHON

According to Tables 6.13, 6.14 and 6.15 Python used less than expected in our forecast. Especially drop in Quarterly Stack Overflow data is significant but Monthly Stack Overflow and GitHub data show drop is not so critical.

| StackOverflow Python Actual-Prediction (Quarterly) (% Error) | | | |
|---|---|---|---|
| | Actual | Prediction | % Error |
| 2014 Q1 | 31963 | 30057.62 | 5.96% |
| 2014 Q2 | 31305 | 32484.37 | 3.77% |
| 2014 Q3 | 30108 | 34911.13 | 15.95% |
| 2014 Q4 | 30833 | 37337.89 | 21.10% |

***Table 6.13:*** *Quarterly Relative error of Python Prediction on StackExchange Data*

$$MRE = \frac{1}{N}\sum_{1}^{N}\frac{A_i - P_i}{A_i} = \%11.69$$

$$MdRE = \%10.96$$

| StackOverflow Python Actual-Prediction (Monthly) (% Error) | | | |
|---|---|---|---|
| | Actual | Prediction | % Error |
| Jan-14 | 9964 | 10139.27 | 1.76% |
| Feb-14 | 10012 | 9927.45 | 0.84% |
| Mar-14 | 11987 | 11077.73 | 7.59% |
| Apr-14 | 11676 | 10272.86 | 12.02% |
| May-14 | 10012 | 10865.26 | 8.52% |
| Jun-14 | 9617 | 10910.58 | 13.45% |
| Jul-14 | 10715 | 11938.32 | 11.42% |
| Aug-14 | 9820 | 11362.51 | 15.71% |
| Sep-14 | 9573 | 10457.79 | 9.24% |
| Oct-14 | 10513 | 11835.67 | 12.58% |
| Nov-14 | 10532 | 12012.39 | 14.06% |
| Dec-14 | 9788 | 11006.02 | 12.44% |
| Jan-15 | 10547 | 12803.28 | 21.39% |
| Feb-15 | 11495 | 12479.92 | 8.57% |

***Table 6.14:*** *Monthly Relative error of Python Prediction on StackExchange Data*

$$MRE = \frac{1}{N}\sum_{1}^{N}\frac{A_i - P_i}{A_i} = \%10.68$$

$$MdRE = \%12.23$$

| GitHub Python Actual-Prediction (Monthly) (% Error) | | | |
|---|---|---|---|
| | Actual | Prediction | % Error |
| Jan-14 | 767315 | 637537.50 | 16.91% |
| Feb-14 | 892922 | 767302.00 | 14.07% |
| Mar-14 | 900103 | 892909.40 | 0.80% |
| Apr-14 | 903274 | 900102.30 | 0.35% |
| May-14 | 928777 | 903273.70 | 2.75% |
| Jun-14 | 904522 | 928774.40 | 2.68% |
| Jul-14 | 1016891 | 904524.40 | 11.05% |
| Aug-14 | 1035691 | 1016880.00 | 1.82% |
| Sep-14 | 1047753 | 1035689.00 | 1.15% |
| Oct-14 | 1108261 | 1090211.00 | 1.63% |
| Nov-14 | 1065634 | 1151742.00 | 8.08% |
| Dec-14 | 1037617 | 1104693.00 | 6.46% |

**Table 6.15:** Quarterly Relative error of Python Prediction on StackExchange Data

$$MRE = \frac{1}{N}\sum_{1}^{N}\frac{A_i - P_i}{A_i} = \%5.65$$

$$MdRE = \%2.71$$

## 1.6. C++

All three of the forecasts in Tables 6.16, 6.17 and 6.18 shows C++ mentioned less than expected than in forecasts.

| StackOverflow C++ Actual-Prediction (Quarterly) (% Error) | | | |
|---|---|---|---|
| | Actual | Prediction | % Error |
| 2014 Q1 | 24768 | 22756.03 | 8.12% |
| 2014 Q2 | 22159 | 23764.58 | 7.25% |
| 2014 Q3 | 19832 | 24773.13 | 24.91% |
| 2014 Q4 | 21166 | 25781.69 | 21.81% |

***Table 6.16:*** *Quarterly Relative error of C++ Prediction on StackExchange Data*

$$MRE = \frac{1}{N}\sum_{1}^{N}\frac{A_i - P_i}{A_i} = \%15.52$$

$$MdRE = \%14.97$$

| StackOverflow C++ Actual-Prediction (Monthly) (% Error) | | | |
|---|---|---|---|
| | Actual | Prediction | % Error |
| Jan-14 | 7831 | 8136.93 | 3.91% |
| Feb-14 | 7825 | 8315.16 | 6.26% |
| Mar-14 | 9112 | 9126.68 | 0.16% |
| Apr-14 | 8718 | 8748.16 | 0.35% |
| May-14 | 7264 | 8342.32 | 14.84% |
| Jun-14 | 6177 | 8275.93 | 33.98% |
| Jul-14 | 6788 | 8527.37 | 25.62% |
| Aug-14 | 6326 | 8181.51 | 29.33% |
| Sep-14 | 6718 | 8132.54 | 21.06% |
| Oct-14 | 7353 | 9163.18 | 24.62% |
| Nov-14 | 7192 | 9454.68 | 31.46% |
| Dec-14 | 6621 | 8470.07 | 27.93% |
| Jan-15 | 6679 | 9530.55 | 42.69% |
| Feb-15 | 7317 | 9719.26 | 32.83% |

***Table 6.17:*** *Monthly Relative error of C++ Prediction on StackExchange Data*

$$MRE = \frac{1}{N}\sum_{1}^{N}\frac{A_i - P_i}{A_i} = \%21.07$$

$$MdRE = \%26.78$$

| GitHub C++ Actual-Prediction (Monthly) (% Error) | | | |
|---|---|---|---|
| | Actual | Prediction | % Error |
| Jan-14 | 409684 | 384202.30 | 6.22% |
| Feb-14 | 428745 | 409681.40 | 4.45% |
| Mar-14 | 479865 | 428743.10 | 10.65% |
| Apr-14 | 483783 | 479859.90 | 0.81% |
| May-14 | 519419 | 483782.60 | 6.86% |
| Jun-14 | 480760 | 519415.40 | 8.04% |
| Jul-14 | 515337 | 480763.90 | 6.71% |
| Aug-14 | 509015 | 515333.50 | 1.24% |
| Sep-14 | 519585 | 509015.60 | 2.03% |
| Oct-14 | 621810 | 519583.90 | 16.44% |
| Nov-14 | 625806 | 621799.80 | 0.64% |
| Dec-14 | 615837 | 625805.60 | 1.62% |

**Table 6.18:** Quarterly Relative error of C++ Prediction on StackExchange Data

$$MRE = \frac{1}{N}\sum_{1}^{N}\frac{A_i - P_i}{A_i} = \%5.48$$

$$MdRE = \%5.33$$

## 1.7. SQL

According to Table 6.19 and 6.20 shows significant drop in count of posts mentioned SQL in 4th quarter of 2014. No estimate can be made to this kind of drop. We must reason this result in a future research.

| StackOverflow SQL Actual-Prediction (Quarterly) (% Error) | | | |
|---|---|---|---|
| | Actual | Prediction | % Error |
| 2014 Q1 | 22281 | 20679.81 | 7.19% |
| 2014 Q2 | 22402 | 22354.84 | 0.21% |
| 2014 Q3 | 20680 | 24029.87 | 16.20% |
| 2014 Q4 | 14760 | 25704.90 | 74.15% |

***Table 6.19:*** *Quarterly Relative error of SQL Prediction on StackExchange Data*

$$MRE = \frac{1}{N}\sum_{1}^{N}\frac{A_i - P_i}{A_i} = \%24.44$$

$$MdRE = \%11.69$$

| StackOverflow SQL Actual-Prediction (Monthly) (% Error) | | | |
|---|---|---|---|
| | Actual | Prediction | % Error |
| Jan-14 | 7137 | 6563.03 | 8.04% |
| Feb-14 | 7038 | 6649.03 | 5.53% |
| Mar-14 | 8106 | 6735.02 | 16.91% |
| Apr-14 | 8331 | 6821.01 | 18.12% |
| May-14 | 7243 | 6907.00 | 4.64% |
| Jun-14 | 6828 | 6993.00 | 2.42% |
| Jul-14 | 7308 | 7078.99 | 3.13% |
| Aug-14 | 6612 | 7164.98 | 8.36% |
| Sep-14 | 6760 | 7250.97 | 7.26% |
| Oct-14 | 5803 | 7336.97 | 26.43% |
| Nov-14 | 4526 | 7422.96 | 64.01% |
| Dec-14 | 4431 | 7508.95 | 69.46% |
| Jan-15 | 4705 | 7594.95 | 61.42% |
| Feb-15 | 4761 | 7680.94 | 61.33% |

***Table 6.20:*** *Monthly Relative error of SQL Prediction on StackExchange Data*

$$MRE = \frac{1}{N}\sum_{1}^{N}\frac{A_i - P_i}{A_i} = \%25.51$$

$$MdRE = \%17.52$$

Not enough data present to make prediction in GitHub Dataset

## 1.8. C

Stack Overflow data and its resulting Tables 6.21 and 6.22 shows C is mentioned less than expected in 2014. Table 6.23 shows that GitHub data is also supports this drop but not much has happened in other two.

| Stackoverflow C Actual-Prediction (Quarterly) (% Error) | | | |
|---|---|---|---|
| | Actual | Prediction | % Error |
| 2014 Q1 | 12650 | 12005.35 | 5.10% |
| 2014 Q2 | 10663 | 12235.88 | 14.75% |
| 2014 Q3 | 9287 | 11943.88 | 28.61% |
| 2014 Q4 | 11280 | 14454.81 | 28.15% |

***Table 6.21:*** *Quarterly Relative error of C Prediction on StackExchange Data*

$$MRE = \frac{1}{N}\sum_{1}^{N}\frac{A_i - P_i}{A_i} = \%19.15$$

$$MdRE = \%21.45$$

| StackOverflow C Actual-Prediction (Monthly) (% Error) | | | |
|---|---|---|---|
| | Actual | Prediction | % Error |
| Jan-14 | 3812 | 4061.97 | 6.56% |
| Feb-14 | 4052 | 4133.84 | 2.02% |
| Mar-14 | 4786 | 4684.19 | 2.13% |
| Apr-14 | 4258 | 4597.96 | 7.98% |
| May-14 | 3469 | 4178.78 | 20.46% |
| Jun-14 | 2936 | 3925.27 | 33.69% |
| Jul-14 | 3019 | 3952.55 | 30.92% |
| Aug-14 | 2831 | 3926.36 | 38.69% |
| Sep-14 | 3437 | 4054.71 | 17.97% |
| Oct-14 | 4029 | 4802.16 | 19.19% |
| Nov-14 | 3913 | 5123.93 | 30.95% |
| Dec-14 | 3338 | 4325.61 | 29.59% |
| Jan-15 | 3399 | 4793.02 | 41.01% |
| Feb-15 | 3697 | 4866.83 | 31.64% |

**Table 6.22:** Monthly Relative error of C Prediction on StackExchange Data

$$MRE = \frac{1}{N}\sum_{1}^{N}\frac{A_i - P_i}{A_i} = \%22.34$$

$$MdRE = \%30.25$$

| GitHub C Actual-Prediction (Monthly) (% Error) | | | |
|---|---|---|---|
| | Actual | Prediction | % Error |
| Jan-14 | 399192 | 370480.90 | 7.19% |
| Feb-14 | 420197 | 394018.00 | 6.23% |
| Mar-14 | 491779 | 415886.60 | 15.43% |
| Apr-14 | 467028 | 481614.50 | 3.12% |
| May-14 | 468806 | 469045.50 | 0.05% |
| Jun-14 | 489748 | 468842.10 | 4.27% |
| Jul-14 | 459340 | 486615.80 | 5.94% |
| Aug-14 | 438367 | 463580.00 | 5.75% |
| Sep-14 | 448036 | 442132.50 | 1.32% |
| Oct-14 | 484921 | 447145.50 | 7.79% |
| Nov-14 | 512953 | 479187.20 | 6.58% |
| Dec-14 | 490187 | 508457.80 | 3.73% |

**Table 6.23:** Quarterly Relative error of C Prediction on StackExchange Data

$$MRE = \frac{1}{N}\sum_{1}^{N}\frac{A_i - P_i}{A_i} = \%5.62$$

$$MdRE = \%5.84$$

## 1.9. SWIFT

Not enough data to make prediction in StackOverflow Dataset

Not enough data to make prediction in GitHub Dataset

# 1.10.R

Table 6.24, 6.25 and 6.26 show rise in count of posts that mentions R. This is not a surprise as R slowly became standard in data analysis. As an example even Microsoft is planning to add this language to its SQL Server in 2016 version [28]. All signs shows that R is a language that we may invest on.

| Stackoverflow R Actual-Prediction (Quarterly) (% Error) | | | |
|---|---|---|---|
| | Actual | Prediction | % Error |
| 2014 Q1 | 7965 | 6783.19 | 14.84% |
| 2014 Q2 | 8255 | 7291.78 | 11.67% |
| 2014 Q3 | 8049 | 7800.37 | 3.09% |
| 2014 Q4 | 8547 | 8308.96 | 2.79% |

***Table 6.24:*** *Quarterly Relative error of R Prediction on StackExchange Data*

$$MRE = \frac{1}{N}\sum_{1}^{N}\frac{A_i - P_i}{A_i} = \%8.09$$

$$MdRE = \%7.38$$

| StackOverflow R Actual-Prediction (Monthly) (% Error) | | | |
|---|---|---|---|
| | Actual | Prediction | % Error |
| Jan-14 | 2380 | 2130.26 | 10.49% |
| Feb-14 | 2643 | 2177.73 | 17.60% |
| Mar-14 | 2942 | 2225.20 | 24.36% |
| Apr-14 | 2963 | 2272.67 | 23.30% |
| May-14 | 2751 | 2320.14 | 15.66% |
| Jun-14 | 2541 | 2367.61 | 6.82% |
| Jul-14 | 2853 | 2415.08 | 15.35% |
| Aug-14 | 2647 | 2462.54 | 6.97% |
| Sep-14 | 2549 | 2510.01 | 1.53% |
| Oct-14 | 3001 | 2557.48 | 14.78% |
| Nov-14 | 2997 | 2604.95 | 13.08% |
| Dec-14 | 2549 | 2652.42 | 4.06% |
| Jan-15 | 2902 | 2699.89 | 6.96% |
| Feb-15 | 3361 | 2747.36 | 18.26% |

**Table 6.25:** Monthly Relative error of R Prediction on StackExchange Data

$$MRE = \frac{1}{N}\sum_{1}^{N}\frac{A_i - P_i}{A_i} = \%12.80$$

$$MdRE = \%13.93$$

| GitHub R Actual-Prediction (Monthly) (% Error) | | | |
|---|---|---|---|
| | Actual | Prediction | % Error |
| Jan-14 | 33002 | 23152.09 | 29.85% |
| Feb-14 | 33125 | 33001.01 | 0.37% |
| Mar-14 | 53197 | 33124.99 | 37.73% |
| Apr-14 | 70530 | 53194.98 | 24.58% |
| May-14 | 71445 | 80753.71 | 13.03% |
| Jun-14 | 71542 | 77245.42 | 7.97% |
| Jul-14 | 75875 | 73246.61 | 3.46% |
| Aug-14 | 76262 | 79727.02 | 4.54% |
| Sep-14 | 80267 | 77967.91 | 2.86% |
| Oct-14 | 79211 | 81972.80 | 3.49% |
| Nov-14 | 81287 | 80917.80 | 0.45% |
| Dec-14 | 75988 | 82993.75 | 9.22% |

**Table 6.26:** Monthly Relative error of R Prediction on GitHub Data

$$MRE = \frac{1}{N}\sum_{1}^{N}\frac{A_i - P_i}{A_i} = \%11.46$$

$$MdRE = \%6.26$$

## 1.11.OBJECTIVE-C

As we look Tables 6.27, 6.28 and 6.29 all three show us the drop of usage in 4$^{th}$ quarter of 2014. That is not a surprise as releases of Swift has made huge impact in Mac developer community. This might be a sign of a migration between Mac developers from Objective C to Swift. Because as a relatively constant data even GitHub shows a steep drop at December 2014 following by a higher than expected event count in 2$^{nd}$ half of the 2014. This might be a sign of some projects already starting to move Swift. This might be an interesting topic for a future research.

| Stackoverflow Objective C Actual-Prediction (Quarterly) (% Error) | | | |
|---|---|---|---|
| | Actual | Prediction | % Error |
| 2014 Q1 | 14366 | 12626.88 | 12.11% |
| 2014 Q2 | 13845 | 12626.88 | 8.80% |
| 2014 Q3 | 13081 | 12626.88 | 3.47% |
| 2014 Q4 | 11328 | 12626.88 | 11.47% |

**Table 6.27:** Quarterly Relative error of Objective C Prediction on StackExchange Data

$$MRE = \frac{1}{N}\sum_{1}^{N}\frac{A_i - P_i}{A_i} = \%8.96$$

$$MdRE = \%10.13$$

| StackOverflow Objective C Actual-Prediction (Monthly) (% Error) | | | |
|---|---|---|---|
| | Actual | Prediction | % Error |
| Jan-14 | 4450 | 4176.90 | 6.14% |
| Feb-14 | 4639 | 4254.98 | 8.28% |
| Mar-14 | 5277 | 4333.07 | 17.89% |
| Apr-14 | 5104 | 4411.15 | 13.57% |
| May-14 | 4513 | 4489.23 | 0.53% |
| Jun-14 | 4228 | 4567.31 | 8.03% |
| Jul-14 | 4633 | 4645.39 | 0.27% |
| Aug-14 | 4168 | 4723.47 | 13.33% |
| Sep-14 | 4280 | 4801.55 | 12.19% |
| Oct-14 | 4190 | 4879.63 | 16.46% |
| Nov-14 | 3731 | 4957.71 | 32.88% |
| Dec-14 | 3407 | 5035.79 | 47.81% |
| Jan-15 | 3405 | 5113.87 | 50.19% |
| Feb-15 | 3695 | 5191.95 | 40.51% |

**Table 6.28:** Monthly Relative error of Objective C Prediction on StackExchange Data

$$MRE = \frac{1}{N}\sum_{1}^{N}\frac{A_i - P_i}{A_i} = \%19.15$$

$$MdRE = \%15.02$$

| GitHub Objective C Actual-Prediction (Monthly) (% Error) | | | |
|---|---|---|---|
| | Actual | Prediction | % Error |
| Jan-14 | 199228 | 176967.20 | 11.17% |
| Feb-14 | 198225 | 198284.30 | 0.03% |
| Mar-14 | 241192 | 198227.50 | 17.81% |
| Apr-14 | 240943 | 239427.70 | 0.63% |
| May-14 | 242050 | 240726.50 | 0.55% |
| Jun-14 | 192671 | 241952.20 | 25.58% |
| Jul-14 | 229552 | 196481.30 | 14.41% |
| Aug-14 | 225826 | 227202.80 | 0.61% |
| Sep-14 | 217323 | 225922.40 | 3.96% |
| Oct-14 | 231811 | 218442.90 | 5.77% |
| Nov-14 | 231248 | 230000.10 | 0.54% |
| Dec-14 | 211675 | 231215.60 | 9.23% |

**Table 6.29:** Monthly Relative error of Objective C Prediction on GitHub Data

$$MRE = \frac{1}{N}\sum_{1}^{N}\frac{A_i - P_i}{A_i} = \%7.52$$

$$MdRE = \%4.86$$

## 1.12. VB.NET

Tables 6.30, 6.31 and 6.32 shows drops in usage of Visual Basic with respected to the forecasting values. Those drop reaches nearly %20 in the end of the 2014. After 50 years of its first release Basic is still in use in a respected amount. But as the figures show there are better options than Basic and real applications already moved to those languages.

| Stackoverflow VB.NET Actual-Prediction (Quarterly) (% Error) | | | |
|---|---|---|---|
| | Actual | Prediction | % Error |
| 2014 Q1 | 5751 | 5178.81 | 9.95% |
| 2014 Q2 | 5399 | 5375.81 | 0.43% |
| 2014 Q3 | 4784 | 5572.82 | 16.49% |
| 2014 Q4 | 4687 | 5769.83 | 23.10% |

**Table 6.30:** Quarterly Relative error of VB.NET Prediction on StackExchange Data

$$MRE = \frac{1}{N}\sum_{1}^{N}\frac{A_i - P_i}{A_i} = \%12.49$$

$$MdRE = \%13.22$$

| StackOverflow VB.NET Actual-Prediction (Monthly) (% Error) | | | |
|---|---|---|---|
| | Actual | Prediction | % Error |
| Jan-14 | 1831 | 1708.97 | 6.66% |
| Feb-14 | 1761 | 1733.42 | 1.57% |
| Mar-14 | 2159 | 1757.87 | 18.58% |
| Apr-14 | 2130 | 1782.32 | 16.32% |
| May-14 | 1696 | 1806.77 | 6.53% |
| Jun-14 | 1573 | 1831.23 | 16.42% |
| Jul-14 | 1731 | 1855.68 | 7.20% |
| Aug-14 | 1510 | 1880.13 | 24.51% |
| Sep-14 | 1543 | 1904.58 | 23.43% |
| Oct-14 | 1632 | 1929.03 | 18.20% |
| Nov-14 | 1539 | 1953.48 | 26.93% |
| Dec-14 | 1516 | 1977.93 | 30.47% |
| Jan-15 | 1481 | 2002.38 | 35.20% |
| Feb-15 | 1567 | 2026.83 | 29.34% |

**Table 6.31:** Monthly Relative error of VB.NET Prediction on StackExchange Data

$$MRE = \frac{1}{N}\sum_{1}^{N}\frac{A_i - P_i}{A_i} = \%19.15$$

$$MdRE = \%21.01$$

| GitHub VB.NET Actual-Prediction (Monthly) (% Error) | | | |
|---|---|---|---|
| | Actual | Prediction | % Error |
| Jan-14 | 5301 | 4470.09 | 15.67% |
| Feb-14 | 5581 | 5266.85 | 5.63% |
| Mar-14 | 6180 | 5571.85 | 9.84% |
| Apr-14 | 7019 | 6167.23 | 12.14% |
| May-14 | 7346 | 7018.86 | 4.45% |
| Jun-14 | 7396 | 7345.97 | 0.68% |
| Jul-14 | 6882 | 7396.00 | 7.47% |
| Aug-14 | 5966 | 6882.05 | 15.35% |
| Sep-14 | 7573 | 5966.09 | 21.22% |
| Oct-14 | 10281 | 7526.95 | 26.79% |
| Nov-14 | 10115 | 10280.73 | 1.64% |
| Dec-14 | 8429 | 10115.02 | 20.00% |

**Table 6.32:** Monthly Relative error of VB.NET Prediction on GitHub Data

$$MRE = \frac{1}{N}\sum_{1}^{N}\frac{A_i - P_i}{A_i} = \%11.74$$

$$MdRE = \%10.99$$

## 1.13.MATLAB

All three table (Table 6.33, 6.34 and 6.35) shows inconsistent results with forecasts. This might be related with the seasonality of the data which causes peak point in some time frames. Still not much lower than expectations at least in GitHub data. As a last point our data shows Matlab is still most used numerical analysis language.

| Stackoverflow Matlab Actual-Prediction (Quarterly) (% Error) | | | |
|---|---|---|---|
| | Actual | Prediction | % Error |
| 2014 Q1 | 3954 | 3561.82 | 9.92% |
| 2014 Q2 | 4189 | 3876.87 | 7.45% |
| 2014 Q3 | 3381 | 4195.71 | 24.10% |
| 2014 Q4 | 3895 | 4516.61 | 15.96% |

***Table 6.33:*** *Quarterly Relative error of MATLAB Prediction on StackExchange Data*

$$MRE = \frac{1}{N}\sum_{1}^{N}\frac{A_i - P_i}{A_i} = \%14.36$$
$$MdRE = \%13.75$$

| StackOverflow Matlab Actual-Prediction (Monthly) (% Error) | | | |
|---|---|---|---|
| | Actual | Prediction | % Error |
| Jan-14 | 1145 | 1084.71 | 5.27% |
| Feb-14 | 1265 | 1092.41 | 13.64% |
| Mar-14 | 1544 | 1100.11 | 28.75% |
| Apr-14 | 1602 | 1107.81 | 30.85% |
| May-14 | 1445 | 1115.50 | 22.80% |
| Jun-14 | 1142 | 1123.20 | 1.65% |
| Jul-14 | 1201 | 1130.90 | 5.84% |
| Aug-14 | 1006 | 1138.60 | 13.18% |
| Sep-14 | 1174 | 1146.30 | 2.36% |
| Oct-14 | 1347 | 1153.99 | 14.33% |
| Nov-14 | 1415 | 1161.69 | 17.90% |
| Dec-14 | 1133 | 1169.39 | 3.21% |
| Jan-15 | 1123 | 1177.09 | 4.82% |
| Feb-15 | 1404 | 1184.78 | 15.61% |

**Table 6.34:** Monthly Relative error of MATLAB Prediction on StackExchange Data

$$MRE = \frac{1}{N}\sum_{1}^{N}\frac{A_i - P_i}{A_i} = \%12.87$$
$$MdRE = \%13.75$$

| GitHub Matlab Actual-Prediction (Monthly) (% Error) | | | |
|---|---|---|---|
| | Actual | Prediction | % Error |
| Jan-14 | 13148 | 15966.14 | 21.43% |
| Feb-14 | 13462 | 13148.28 | 2.33% |
| Mar-14 | 15956 | 13461.97 | 15.63% |
| Apr-14 | 17148 | 15955.75 | 6.95% |
| May-14 | 18055 | 17067.38 | 5.47% |
| Jun-14 | 18519 | 18054.91 | 2.51% |
| Jul-14 | 17899 | 18518.95 | 3.46% |
| Aug-14 | 16911 | 17899.06 | 5.84% |
| Sep-14 | 18761 | 16911.10 | 9.86% |
| Oct-14 | 24243 | 18760.81 | 22.61% |
| Nov-14 | 27570 | 24242.45 | 12.07% |
| Dec-14 | 23421 | 27569.67 | 17.71% |

**Table 6.35:** Monthly Relative error of MATLAB Prediction on GitHub Data

$$MRE = \frac{1}{N}\sum_{1}^{N}\frac{A_i - P_i}{A_i} = \%10.49$$

$$MdRE = \%8.41$$

## 1.14.SCALA

Tables 6.36, 6.37 and 6.38 shows usage of Matlab is harmonious with our forecasts. We can say that as a prominent functional language Scala maintain its market share among the programming languages.

| Stackoverflow Scala Actual-Prediction (Quarterly) (% Error) | | | |
|---|---|---|---|
| | Actual | Prediction | % Error |
| 2014 Q1 | 2563 | 2575.39 | 0.48% |
| 2014 Q2 | 2824 | 2677.04 | 5.20% |
| 2014 Q3 | 2754 | 2778.69 | 0.90% |
| 2014 Q4 | 2741 | 2880.34 | 5.08% |

***Table 6.36:*** *Quarterly Relative error of Scala Prediction on StackExchange Data*

$$MRE = \frac{1}{N}\sum_{1}^{N}\frac{A_i - P_i}{A_i} = \%2.92$$

$$MdRE = \%2.99$$

| StackOverflow Scala Actual-Prediction (Monthly) (% Error) | | | |
|---|---|---|---|
| | Actual | Prediction | % Error |
| Jan-14 | 775 | 849.81 | 9.65% |
| Feb-14 | 834 | 863.93 | 3.59% |
| Mar-14 | 954 | 878.05 | 7.96% |
| Apr-14 | 971 | 892.17 | 8.12% |
| May-14 | 878 | 906.29 | 3.22% |
| Jun-14 | 975 | 920.41 | 5.60% |
| Jul-14 | 948 | 934.53 | 1.42% |
| Aug-14 | 906 | 948.65 | 4.71% |
| Sep-14 | 900 | 962.77 | 6.97% |
| Oct-14 | 900 | 976.89 | 8.54% |
| Nov-14 | 946 | 991.01 | 4.76% |
| Dec-14 | 895 | 1005.13 | 12.30% |
| Jan-15 | 944 | 1019.25 | 7.97% |
| Feb-15 | 939 | 1033.36 | 10.05% |

**Table 6.37:** Monthly Relative error of Scala Prediction on StackExchange Data

$$MRE = \frac{1}{N}\sum_{1}^{N}\frac{A_i - P_i}{A_i} = \%6.78$$

$$MdRE = \%7.47$$

| GitHub Scala Actual-Prediction (Monthly) (% Error) | | | |
|---|---|---|---|
| | Actual | Prediction | % Error |
| Jan-14 | 62318 | 52603.79 | 15.59% |
| Feb-14 | 67919 | 60537.15 | 10.87% |
| Mar-14 | 74460 | 65736.13 | 11.72% |
| Apr-14 | 77626 | 73104.16 | 5.83% |
| May-14 | 82026 | 76873.60 | 6.28% |
| Jun-14 | 82092 | 81174.15 | 1.12% |
| Jul-14 | 91327 | 81941.81 | 10.28% |
| Aug-14 | 86946 | 89078.34 | 2.45% |
| Sep-14 | 89033 | 87414.90 | 1.82% |
| Oct-14 | 98360 | 88764.37 | 9.76% |
| Nov-14 | 98141 | 96777.12 | 1.39% |
| Dec-14 | 93248 | 97922.62 | 5.01% |

**Table 6.38:** Monthly Relative error of Scala Prediction on GitHub Data

$$MRE = \frac{1}{N}\sum_{1}^{N}\frac{A_i - P_i}{A_i} = \%6.84$$

$$MdRE = \%2.71$$

## 1.15.PERL

Tables 6.39, 6.40 and 6.41 shows the obvious decline of PERL in numbers. This decline is because of the new languages or technologies already better than PERL available in many context. For example R in data processing, node.js in CGI development, Python in nearly every aspect could preferable among the developers. For more information an article by Conor Myhrvold [29] is a good reading about the decline of the Perl.

| Stackoverflow Perl Actual-Prediction (Quarterly) (% Error) | | | |
|---|---|---|---|
| | Actual | Prediction | % Error |
| 2014 Q1 | 2631 | 2686.66 | 2.12% |
| 2014 Q2 | 2435 | 2769.25 | 13.73% |
| 2014 Q3 | 2239 | 2849.64 | 27.27% |
| 2014 Q4 | 1983 | 2927.88 | 47.65% |

***Table 6.39:** Quarterly Relative error of PERL Prediction on StackExchange Data*

$$MRE = \frac{1}{N}\sum_{1}^{N}\frac{A_i - P_i}{A_i} = \%22.69$$

$$MdRE = \%20.50$$

| StackOverflow Perl Actual-Prediction (Monthly) (% Error) | | | |
|---|---|---|---|
| | Actual | Prediction | % Error |
| Jan-14 | 840 | 864.39 | 2.90% |
| Feb-14 | 828 | 879.11 | 6.17% |
| Mar-14 | 963 | 893.82 | 7.18% |
| Apr-14 | 868 | 908.54 | 4.67% |
| May-14 | 843 | 923.26 | 9.52% |
| Jun-14 | 724 | 937.98 | 29.55% |
| Jul-14 | 789 | 952.69 | 20.75% |
| Aug-14 | 731 | 967.41 | 32.34% |
| Sep-14 | 719 | 982.13 | 36.60% |
| Oct-14 | 684 | 996.85 | 45.74% |
| Nov-14 | 695 | 1011.56 | 45.55% |
| Dec-14 | 604 | 1026.28 | 69.91% |
| Jan-15 | 635 | 1041.00 | 63.94% |
| Feb-15 | 607 | 1055.72 | 73.92% |

**Table 6.40:** Monthly Relative error of PERL Prediction on StackExchange Data

$$MRE = \frac{1}{N}\sum_{1}^{N}\frac{A_i - P_i}{A_i} = \%32.05$$

$$MdRE = \%34.47$$

| GitHub Perl Actual-Prediction (Monthly) (% Error) | | | |
|---|---|---|---|
| | Actual | Prediction | % Error |
| Jan-14 | 66393 | 81451.72 | 22.68% |
| Feb-14 | 58988 | 79700.29 | 35.11% |
| Mar-14 | 69935 | 77351.35 | 10.60% |
| Apr-14 | 65176 | 76537.16 | 17.43% |
| May-14 | 68727 | 75428.34 | 9.75% |
| Jun-14 | 66044 | 74714.35 | 13.13% |
| Jul-14 | 73340 | 73891.63 | 0.75% |
| Aug-14 | 71185 | 73806.38 | 3.68% |
| Sep-14 | 75194 | 73594.32 | 2.13% |
| Oct-14 | 119219 | 73740.54 | 38.15% |
| Nov-14 | 76933 | 77306.60 | 0.49% |
| Dec-14 | 69633 | 77312.33 | 11.03% |

**Table 6.41:** Monthly Relative error of PERL Prediction on GitHub Data

$$MRE = \frac{1}{N}\sum_{1}^{N}\frac{A_i - P_i}{A_i} = \%13.74$$

$$MdRE = \%10.82$$

## 1.16. HASKELL

All three forecast on Tables 6.42, 6.43 and 6.44 shows increase in usage of Haskell. This increase is not surprise as being a prominent functional language. For an add-on as a most used “pure functional language” its increased usage will be a sign of a future paradigm shift in programming languages. This sign of a possible paradigm shift might be a good starting point for a future research.

| Stackoverflow Haskell Actual-Prediction (Quarterly) (% Error) | | | |
|---|---|---|---|
| | Actual | Prediction | % Error |
| 2014 Q1 | 1512 | 1355.22 | 10.37% |
| 2014 Q2 | 1507 | 1403.57 | 6.86% |
| 2014 Q3 | 1332 | 1447.63 | 8.68% |
| 2014 Q4 | 1705 | 1487.63 | 12.75% |

***Table 6.42:*** *Quarterly Relative error of Haskell Prediction on StackExchange Data*

$$MRE = \frac{1}{N}\sum_{1}^{N}\frac{A_i - P_i}{A_i} = \%9.67$$

$$MdRE = \%9.52$$

| StackOverflow Haskell Actual-Prediction (Monthly) (% Error) | | | |
|---|---|---|---|
| | Actual | Prediction | % Error |
| Jan-14 | 483 | 462.99 | 4.14% |
| Feb-14 | 477 | 470.76 | 1.31% |
| Mar-14 | 552 | 478.53 | 13.31% |
| Apr-14 | 513 | 486.30 | 5.20% |
| May-14 | 562 | 494.07 | 12.09% |
| Jun-14 | 432 | 501.84 | 16.17% |
| Jul-14 | 473 | 509.61 | 7.74% |
| Aug-14 | 435 | 517.38 | 18.94% |
| Sep-14 | 424 | 525.15 | 23.86% |
| Oct-14 | 558 | 532.92 | 4.50% |
| Nov-14 | 601 | 540.69 | 10.04% |
| Dec-14 | 546 | 548.46 | 0.45% |
| Jan-15 | 504 | 556.22 | 10.36% |
| Feb-15 | 425 | 563.99 | 32.70% |

**Table 6.43:** Monthly Relative error of Haskell Prediction on StackExchange Data

$$MRE = \frac{1}{N}\sum_{1}^{N}\frac{A_i - P_i}{A_i} = \%11.49$$

$$MdRE = \%11.22$$

| GitHub Haskell Actual-Prediction (Monthly) (% Error) | | | |
|---|---|---|---|
| | Actual | Prediction | % Error |
| Jan-14 | 32630 | 26746.00 | 18.03% |
| Feb-14 | 30053 | 32629.40 | 8.57% |
| Mar-14 | 36710 | 30053.27 | 18.13% |
| Apr-14 | 40305 | 36709.33 | 8.92% |
| May-14 | 41590 | 40304.63 | 3.09% |
| Jun-14 | 42079 | 41589.87 | 1.16% |
| Jul-14 | 40944 | 42078.95 | 2.77% |
| Aug-14 | 42585 | 40944.12 | 3.85% |
| Sep-14 | 40031 | 42584.83 | 6.38% |
| Oct-14 | 45455 | 40031.26 | 11.93% |
| Nov-14 | 43082 | 45454.45 | 5.51% |
| Dec-14 | 46535 | 43082.25 | 7.42% |

**Table 6.44:** Monthly Relative error of Haskell Prediction on GitHub Data

$$MRE = \frac{1}{N}\sum_{1}^{N}\frac{A_i - P_i}{A_i} = \%7.98$$

$$MdRE = \%6.90$$

## 1.17.GO

Tables 6.45, 6.46 and 6.47 shows significant rise of Go usage. As a language supported by Google this increase is not a surprise. Go is a good language to invest especially for the developers which wants to work in Google in the future.

| Stackoverflow Go Actual-Prediction (Quarterly) (% Error) | | | |
|---|---|---|---|
| | Actual | Prediction | % Error |
| 2014 Q1 | 769 | 673.57 | 12.41% |
| 2014 Q2 | 920 | 689.82 | 25.02% |
| 2014 Q3 | 1265 | 706.07 | 44.18% |
| 2014 Q4 | 1098 | 722.32 | 34.22% |

***Table 6.45:*** *Quarterly Relative error of Go Prediction on StackExchange Data*

$$MRE = \frac{1}{N}\sum_{1}^{N}\frac{A_i - P_i}{A_i} = \%28.96$$

$$MdRE = \%29.62$$

| StackOverflow Go Actual-Prediction (Monthly) (% Error) | | | |
|---|---|---|---|
| | Actual | Prediction | % Error |
| Jan-14 | 304 | 243.79 | 19.81% |
| Feb-14 | 209 | 252.29 | 20.71% |
| Mar-14 | 256 | 260.79 | 1.87% |
| Apr-14 | 346 | 269.29 | 22.17% |
| May-14 | 283 | 277.79 | 1.84% |
| Jun-14 | 291 | 286.28 | 1.62% |
| Jul-14 | 463 | 294.78 | 36.33% |
| Aug-14 | 407 | 303.28 | 25.48% |
| Sep-14 | 395 | 311.78 | 21.07% |
| Oct-14 | 391 | 320.28 | 18.09% |
| Nov-14 | 348 | 328.77 | 5.52% |
| Dec-14 | 359 | 337.27 | 6.05% |
| Jan-15 | 496 | 345.77 | 30.29% |
| Feb-15 | 402 | 354.27 | 11.87% |

**Table 6.46:** Monthly Relative error of Go Prediction on StackExchange Data

$$MRE = \frac{1}{N}\sum_{1}^{N}\frac{A_i - P_i}{A_i} = \%15.91$$

$$MdRE = \%14.98$$

| GitHub Go Actual-Prediction (Monthly) (% Error) | | | |
|---|---|---|---|
| | Actual | Prediction | % Error |
| Jan-14 | 86976 | 73629.79 | 15.34% |
| Feb-14 | 87744 | 86974.66 | 0.88% |
| Mar-14 | 92445 | 87743.92 | 5.09% |
| Apr-14 | 107396 | 92444.53 | 13.92% |
| May-14 | 119711 | 108382.50 | 9.46% |
| Jun-14 | 130499 | 124490.70 | 4.60% |
| Jul-14 | 158225 | 137235.80 | 13.27% |
| Aug-14 | 157697 | 167136.80 | 5.99% |
| Sep-14 | 162321 | 170750.30 | 5.19% |
| Oct-14 | 174917 | 172378.60 | 1.45% |
| Nov-14 | 178616 | 183834.20 | 2.92% |
| Dec-14 | 201895 | 188125.60 | 6.82% |

**Table 6.47:** Monthly Relative error of Go Prediction on GitHub Data

$$MRE = \frac{1}{N}\sum_{1}^{N}\frac{A_i - P_i}{A_i} = \%7.08$$

$$MdRE = \%5.59$$

## 1.18.ASSEMBLY

All 3 tables (Tables 6.48, 6.49 and 6.50) shows decreasing usage of assembly with respect to our forecast. As being a low level language this result is not surprising. As technology advances, implemented standards causes less and less requirement for low level programming. Those decrease in numbers are supports our assertion of drop in requirement for low level programming.

| Stackoverflow Assembly Actual-Prediction (Quarterly) (% Error) | | | |
|---|---|---|---|
| | Actual | Prediction | % Error |
| 2014 Q1 | 1193 | 1228.32 | 2.96% |
| 2014 Q2 | 1027 | 1261.70 | 22.85% |
| 2014 Q3 | 734 | 1073.61 | 46.27% |
| 2014 Q4 | 1193 | 1441.95 | 20.87% |

***Table 6.48:*** *Quarterly Relative error of Assembly Prediction on StackExchange Data*

$$MRE = \frac{1}{N}\sum_{1}^{N}\frac{A_i - P_i}{A_i} = \%23.24$$

$$MdRE = \%21.86$$

| StackOverflow Assembly Actual-Prediction (Monthly) (% Error) | | | |
|---|---|---|---|
| | Actual | Prediction | % Error |
| Jan-14 | 344 | 365.87 | 6.36% |
| Feb-14 | 385 | 395.61 | 2.76% |
| Mar-14 | 464 | 456.52 | 1.61% |
| Apr-14 | 396 | 450.00 | 13.64% |
| May-14 | 349 | 396.72 | 13.67% |
| Jun-14 | 282 | 352.16 | 24.88% |
| Jul-14 | 229 | 330.56 | 44.35% |
| Aug-14 | 192 | 315.06 | 64.09% |
| Sep-14 | 313 | 369.09 | 17.92% |
| Oct-14 | 417 | 472.33 | 13.27% |
| Nov-14 | 419 | 499.72 | 19.26% |
| Dec-14 | 357 | 411.68 | 15.32% |
| Jan-15 | 346 | 431.70 | 24.77% |
| Feb-15 | 431 | 465.73 | 8.06% |

**Table 6.49:** Monthly Relative error of Assembly Prediction on StackExchange Data

$$MRE = \frac{1}{N}\sum_{1}^{N}\frac{A_i - P_i}{A_i} = \%19.28$$

$$MdRE = \%16.62$$

| GitHub Assembly Actual-Prediction (Monthly) (% Error) | | | |
|---|---|---|---|
| | Actual | Prediction | % Error |
| Jan-14 | 5520 | 3784.20 | 31.45% |
| Feb-14 | 6719 | 3850.64 | 42.69% |
| Mar-14 | 9024 | 3965.09 | 56.06% |
| Apr-14 | 7015 | 4151.94 | 40.81% |
| May-14 | 8021 | 4247.99 | 47.04% |
| Jun-14 | 5928 | 7836.26 | 32.19% |
| Jul-14 | 16211 | 6488.03 | 59.98% |
| Aug-14 | 11168 | 11795.49 | 5.62% |
| Sep-14 | 10586 | 10927.51 | 3.23% |
| Oct-14 | 14024 | 10937.24 | 22.01% |
| Nov-14 | 12332 | 12514.12 | 1.48% |
| Dec-14 | 8473 | 12417.26 | 46.55% |

**Table 6.50:** Monthly Relative error of Assembly Prediction on GitHub Data

$$MRE = \frac{1}{N}\sum_{1}^{N}\frac{A_i - P_i}{A_i} = \%32.43$$

$$MdRE = \%36.50$$

## 1.19.XSLT

Usage of XSLT is losing its importance as JSON replacing XML as a standard data transfer format. Our assertion about JSON and XML formats are supported by both of our forecast tables. Huge drops in count of posts mentions XSLT shows less and less usage every day.

| Stackoverflow Xslt Actual-Prediction (Quarterly) (% Error) | | | |
|---|---|---|---|
| | Actual | Prediction | % Error |
| 2014 Q1 | 1265 | 1261.11 | 0.31% |
| 2014 Q2 | 1144 | 1303.31 | 13.93% |
| 2014 Q3 | 964 | 1344.67 | 39.49% |
| 2014 Q4 | 1070 | 1385.20 | 29.46% |

***Table 6.51:*** *Quarterly Relative error of XSLT Prediction on StackExchange Data*

$$MRE = \frac{1}{N}\sum_{1}^{N}\frac{A_i - P_i}{A_i} = \%20.79$$

$$MdRE = \%21.69$$

| StackOverflow Xslt Actual-Prediction (Monthly) (% Error) | | | |
|---|---|---|---|
| | Actual | Prediction | % Error |
| Jan-14 | 422 | 431.53 | 2.26% |
| Feb-14 | 424 | 437.85 | 3.27% |
| Mar-14 | 419 | 444.16 | 6.01% |
| Apr-14 | 382 | 450.48 | 17.93% |
| May-14 | 412 | 456.80 | 10.87% |
| Jun-14 | 350 | 463.12 | 32.32% |
| Jul-14 | 354 | 469.44 | 32.61% |
| Aug-14 | 296 | 475.76 | 60.73% |
| Sep-14 | 314 | 482.08 | 53.53% |
| Oct-14 | 329 | 488.40 | 48.45% |
| Nov-14 | 379 | 494.71 | 30.53% |
| Dec-14 | 362 | 501.03 | 38.41% |
| Jan-15 | 325 | 507.35 | 56.11% |
| Feb-15 | 351 | 513.67 | 46.35% |

**Table 6.52:** Monthly Relative error of XSLT Prediction on StackExchange Data

$$MRE = \frac{1}{N}\sum_{1}^{N}\frac{A_i - P_i}{A_i} = \%31.38$$

$$MdRE = \%35.51$$

Not enough data to make prediction in GitHub Dataset

## 1.20.RUBY

Even Quarterly Forecast data that resides in Table 6.53 show a drop in usage other two forecasts shows harmonious run with actual data. Results are shows Ruby usage is still increasing.

| Stackoverflow Ruby Actual-Prediction (Quarterly) (% Error) | | | |
|---|---|---|---|
| | Actual | Prediction | % Error |
| 2014 Q1 | 26653 | 26397.72 | 0.96% |
| 2014 Q2 | 26693 | 27716.03 | 3.83% |
| 2014 Q3 | 24683 | 29034.33 | 17.63% |
| 2014 Q4 | 22023 | 30352.63 | 37.82% |

***Table 6.53:*** *Quarterly Relative error of Ruby Prediction on StackExchange Data*

$$MRE = \frac{1}{N}\sum_{1}^{N}\frac{A_i - P_i}{A_i} = \%15.06$$

$$MdRE = \%10.73$$

| StackOverflow Ruby Actual-Prediction (Monthly) (% Error) | | | |
|---|---|---|---|
| | Actual | Prediction | % Error |
| Jan-14 | 8797 | 8170.51 | 7.12% |
| Feb-14 | 8247 | 8171.62 | 0.91% |
| Mar-14 | 9609 | 8172.74 | 14.95% |
| Apr-14 | 9427 | 8173.85 | 13.29% |
| May-14 | 9055 | 8174.96 | 9.72% |
| Jun-14 | 8211 | 8176.08 | 0.43% |
| Jul-14 | 8764 | 8177.19 | 6.70% |
| Aug-14 | 8392 | 8178.30 | 2.55% |
| Sep-14 | 7527 | 8179.42 | 8.67% |
| Oct-14 | 7626 | 8180.53 | 7.27% |
| Nov-14 | 7257 | 8181.65 | 12.74% |
| Dec-14 | 7140 | 8182.76 | 14.60% |
| Jan-15 | 7724 | 8183.87 | 5.95% |
| Feb-15 | 7587 | 8184.99 | 7.88% |

**Table 6.54:** Monthly Relative error of Ruby Prediction on StackExchange Data

$$MRE = \frac{1}{N}\sum_{1}^{N}\frac{A_i - P_i}{A_i} = \%8.06$$

$$MdRE = \%8.27$$

| GitHub Ruby Actual-Prediction (Monthly) (% Error) | | | |
|---|---|---|---|
| | Actual | Prediction | % Error |
| Jan-14 | 693342 | 624966.50 | 9.86% |
| Feb-14 | 688583 | 693335.10 | 0.69% |
| Mar-14 | 749703 | 688583.50 | 8.15% |
| Apr-14 | 776983 | 749696.80 | 3.51% |
| May-14 | 773934 | 776980.20 | 0.39% |
| Jun-14 | 713954 | 773934.30 | 8.40% |
| Jul-14 | 751158 | 713960.10 | 4.95% |
| Aug-14 | 752705 | 751154.20 | 0.21% |
| Sep-14 | 756156 | 752704.80 | 0.46% |
| Oct-14 | 793450 | 756155.60 | 4.70% |
| Nov-14 | 761795 | 793446.20 | 4.15% |
| Dec-14 | 738799 | 761798.20 | 3.11% |

**Table 6.55:** Monthly Relative error of Ruby Prediction on GitHub Data

$$MRE = \frac{1}{N}\sum_{1}^{N}\frac{A_i - P_i}{A_i} = \%4.05$$

$$MdRE = \%3.83$$

# 2. DATABASES

## 2.1. MICROSOFT SQL SERVER

Tables 6.56 and 6.57 shows a little drop in count of posts that mentions the Microsoft SQL Server in StackExchange. Rise of NoSQL databases may be main reason of this drop. But real answer to this drop may only could be given by a more detailed future research.

| StackExchange SQL Server Actual-Prediction (Quarterly) (% Error) | | | |
|---|---|---|---|
| | Actual | Prediction | % Error |
| 2014 Q1 | 13733 | 13766.76 | 0.25% |
| 2014 Q2 | 13562 | 14322.49 | 5.61% |
| 2014 Q3 | 13365 | 14878.22 | 11.32% |
| 2014 Q4 | 12966 | 15433.96 | 19.03% |

**Table 6.56:** Quarterly Relative error of Microsoft SQL Server Prediction on StackExchange Data

$$MRE = \frac{1}{N}\sum_{1}^{N}\frac{A_i - P_i}{A_i} = \%9.05$$

$$MdRE = \%8.46$$

| StackExchange SQL Server Actual-Prediction (Monthly) (% Error) | | | |
|---|---|---|---|
| | Actual | Prediction | % Error |
| Jan-14 | 4583 | 4448.63 | 2.93% |
| Feb-14 | 4347 | 4509.92 | 3.75% |
| Mar-14 | 4803 | 4571.21 | 4.83% |
| Apr-14 | 4917 | 4632.49 | 5.79% |
| May-14 | 4361 | 4693.78 | 7.63% |
| Jun-14 | 4284 | 4755.06 | 11.00% |
| Jul-14 | 4626 | 4816.35 | 4.11% |
| Aug-14 | 4406 | 4877.63 | 10.70% |
| Sep-14 | 4333 | 4938.92 | 13.98% |
| Oct-14 | 4366 | 5000.20 | 14.53% |
| Nov-14 | 4506 | 5061.49 | 12.33% |
| Dec-14 | 4094 | 5122.78 | 25.13% |
| Jan-15 | 4491 | 5184.06 | 15.43% |
| Feb-15 | 4594 | 5245.35 | 14.18% |

**Table 6.57:** Monthly Relative error of Microsoft SQL Server Prediction on StackExchange Data

$$MRE = \frac{1}{N}\sum_{1}^{N}\frac{A_i - P_i}{A_i} = \%10.45$$

$$MdRE = \%11.66$$

## 2.2. MYSQL

Tables 6.58 and 6.59 shows a little drop in count of posts that mentions the MySQL in StackExchange is very similar to Microsoft SQL Server. Again rise of NoSQL databases may be main reason of this drop. And again real answer to this drop may only could be given by a more detailed future research.

| StackExchange MySQL Actual-Prediction (Quarterly) (% Error) | | | |
|---|---|---|---|
| | Actual | Prediction | % Error |
| 2014 Q1 | 25846 | 23242.55 | 10.07% |
| 2014 Q2 | 24672 | 24362.97 | 1.25% |
| 2014 Q3 | 22209 | 25483.39 | 14.74% |
| 2014 Q4 | 22572 | 26603.82 | 17.86% |

***Table 6.58:*** *Quarterly Relative error of MySQL Prediction on StackExchange Data*

$$MRE = \frac{1}{N}\sum_{1}^{N}\frac{A_i - P_i}{A_i} = \%10.98$$

$$MdRE = \%12.41$$

| StackExchange MySQL Actual-Prediction (Monthly) (% Error) | | | |
|---|---|---|---|
| | Actual | Prediction | % Error |
| Jan-14 | 8253 | 8702.50 | 5.45% |
| Feb-14 | 8121 | 8182.14 | 0.75% |
| Mar-14 | 9472 | 9267.92 | 2.15% |
| Apr-14 | 9376 | 8748.47 | 6.69% |
| May-14 | 8077 | 8444.70 | 4.55% |
| Jun-14 | 7219 | 8695.69 | 20.46% |
| Jul-14 | 7668 | 9230.52 | 20.38% |
| Aug-14 | 7067 | 9450.18 | 33.72% |
| Sep-14 | 7474 | 8635.56 | 15.54% |
| Oct-14 | 7779 | 9009.52 | 15.82% |
| Nov-14 | 7616 | 9159.85 | 20.27% |
| Dec-14 | 7177 | 8707.13 | 21.32% |
| Jan-15 | 7494 | 10553.32 | 40.82% |
| Feb-15 | 7865 | 9891.98 | 25.77% |

***Table 6.59:*** *Monthly Relative error of MySQL Prediction on StackExchange Data*

$$MRE = \frac{1}{N}\sum_{1}^{N}\frac{A_i - P_i}{A_i} = \%16.69$$

$$MdRE = \%20.32$$

## 2.3. ORACLE

Tables 6.60 and 6.61 shows harmonious relation between prediction and actual values of mentions. This result maybe a precursor of success of Oracle’s enterprise market. A future research with detailed data related the other products of the Oracle could give an answer to this results.

| StackExchange Oracle Actual-Prediction (Quarterly) (% Error) | | | |
|---|---|---|---|
| | Actual | Prediction | % Error |
| 2014 Q1 | 5478 | 5266.40 | 3.86% |
| 2014 Q2 | 5842 | 5505.74 | 5.76% |
| 2014 Q3 | 5401 | 5745.08 | 6.37% |
| 2014 Q4 | 5621 | 5984.43 | 6.47% |

***Table 6.60:*** *Quarterly Relative error of Oracle Prediction on StackExchange Data*

$$MRE = \frac{1}{N}\sum_{1}^{N}\frac{A_i - P_i}{A_i} = \%5.61$$

$$MdRE = \%6.06$$

| StackExchange Oracle Actual-Prediction (Monthly) (% Error) | | | |
|---|---|---|---|
| | Actual | Prediction | % Error |
| Jan-14 | 1649 | 1683.89 | 2.12% |
| Feb-14 | 1731 | 1707.71 | 1.35% |
| Mar-14 | 2098 | 1731.53 | 17.47% |
| Apr-14 | 2173 | 1755.35 | 19.22% |
| May-14 | 1939 | 1779.17 | 8.24% |
| Jun-14 | 1730 | 1802.99 | 4.22% |
| Jul-14 | 1819 | 1826.81 | 0.43% |
| Aug-14 | 1658 | 1850.63 | 11.62% |
| Sep-14 | 1924 | 1874.45 | 2.58% |
| Oct-14 | 1853 | 1898.27 | 2.44% |
| Nov-14 | 2017 | 1922.09 | 4.71% |
| Dec-14 | 1751 | 1945.91 | 11.13% |
| Jan-15 | 1719 | 1969.73 | 14.59% |
| Feb-15 | 1942 | 1993.55 | 2.65% |

***Table 6.61:*** *Quarterly Relative error of Oracle Prediction on StackExchange Data*

$$MRE = \frac{1}{N}\sum_{1}^{N}\frac{A_i - P_i}{A_i} = \%7.34$$

$$MdRE = \%6.47$$

## 2.4. MONGODB

Tables 6.62 and 6.63 shows usage of MongoDB is increasing every day. Huge increase in count of posts that mentions MongoDB in StackExchange shows that MongoDB is a good technology to invest for the future.

| StackExchange MongoDB Actual-Prediction (Quarterly) (% Error) | | | |
|---|---|---|---|
| | Actual | Prediction | % Error |
| 2014 Q1 | 3738 | 3105.79 | 16.91% |
| 2014 Q2 | 3994 | 3126.86 | 21.71% |
| 2014 Q3 | 4259 | 3148.08 | 26.08% |
| 2014 Q4 | 4480 | 3169.44 | 29.25% |

***Table 6.62:*** *Quarterly Relative error of MongoDB Prediction on StackExchange Data*

$$MRE = \frac{1}{N}\sum_{1}^{N}\frac{A_i - P_i}{A_i} = \%23.49$$

$$MdRE = \%23.90$$

| StackExchange MongoDB Actual-Prediction (Monthly) (% Error) | | | |
|---|---|---|---|
| | Actual | Prediction | % Error |
| Jan-14 | 1151 | 1005.73 | 12.62% |
| Feb-14 | 1229 | 970.20 | 21.06% |
| Mar-14 | 1358 | 935.94 | 31.08% |
| Apr-14 | 1384 | 902.88 | 34.76% |
| May-14 | 1272 | 870.99 | 31.53% |
| Jun-14 | 1338 | 840.23 | 37.20% |
| Jul-14 | 1458 | 810.56 | 44.41% |
| Aug-14 | 1302 | 781.93 | 39.94% |
| Sep-14 | 1499 | 754.31 | 49.68% |
| Oct-14 | 1515 | 727.67 | 51.97% |
| Nov-14 | 1493 | 701.97 | 52.98% |
| Dec-14 | 1472 | 677.18 | 54.00% |
| Jan-15 | 1531 | 653.26 | 57.33% |
| Feb-15 | 1580 | 630.19 | 60.11% |

***Table 6.63:*** *Monthly Relative error of MongoDB Prediction on StackExchange Data*

$$MRE = \frac{1}{N}\sum_{1}^{N}\frac{A_i - P_i}{A_i} = \%41.33$$

$$MdRE = \%47.04$$

## 2.5. DB2

Tables 6.64 and 6.65 shows DB2 actual count of posts that mentions DB2 harmonious with predicted count of posts that mentions DB2. Again similar with the Oracle being a standard in enterprise market allows DB2 a guaranteed certain popularity.

| StackExchange DB2 Actual-Prediction (Quarterly) (% Error) | | | |
|---|---|---|---|
| | Actual | Prediction | % Error |
| 2014 Q1 | 367 | 373.70 | 1.82% |
| 2014 Q2 | 396 | 394.97 | 0.26% |
| 2014 Q3 | 384 | 415.94 | 8.32% |
| 2014 Q4 | 375 | 436.53 | 16.41% |

***Table 6.64:*** *Quarterly Relative error of DB2 Prediction on StackExchange Data*

$$MRE = \frac{1}{N}\sum_{1}^{N}\frac{A_i - P_i}{A_i} = \%6.70$$

$$MdRE = \%5.07$$

| StackExchange DB2 Actual-Prediction (Monthly) (% Error) | | | |
|---|---|---|---|
| | Actual | Prediction | % Error |
| Jan-14 | 96 | 119.05 | 24.02% |
| Feb-14 | 143 | 120.52 | 15.72% |
| Mar-14 | 128 | 121.99 | 4.69% |
| Apr-14 | 127 | 123.46 | 2.78% |
| May-14 | 121 | 124.93 | 3.25% |
| Jun-14 | 148 | 126.40 | 14.59% |
| Jul-14 | 139 | 127.87 | 8.00% |
| Aug-14 | 131 | 129.34 | 1.26% |
| Sep-14 | 114 | 130.81 | 14.75% |
| Oct-14 | 113 | 132.28 | 17.06% |
| Nov-14 | 125 | 133.75 | 7.00% |
| Dec-14 | 137 | 135.22 | 1.30% |
| Jan-15 | 130 | 136.69 | 5.15% |
| Feb-15 | 137 | 138.16 | 0.85% |

***Table 6.65:*** *Monthly Relative error of DB2 Prediction on StackExchange Data*

$$MRE = \frac{1}{N}\sum_{1}^{N}\frac{A_i - P_i}{A_i} = \%8.60$$

$$MdRE = \%4.92$$

## 2.6. POSTGRESQL

Table 6.66 and 6.67 shows PostgreSQL performs better than expected in forecasts. It has the best performance amongst the RDBM Systems in our research. Even details of this performance must be researched in detailed manner this performance is not a surprise as being open source, having industry standard modules like PostGIS and advantages of scalable architecture makes PostgreSQL as an investable RDBMS solution.

| StackExchange PostgreSQL Actual-Prediction (Quarterly) (% Error) | | | |
|---|---|---|---|
| | Actual | Prediction | % Error |
| 2014 Q1 | 3722 | 3157.15 | 15.18% |
| 2014 Q2 | 3682 | 3266.72 | 11.28% |
| 2014 Q3 | 3685 | 3376.29 | 8.38% |
| 2014 Q4 | 3727 | 3485.87 | 6.47% |

***Table 6.66:** Quarterly Relative error of PostgreSQL Prediction on StackExchange Data*

$$MRE = \frac{1}{N}\sum_{1}^{N}\frac{A_i - P_i}{A_i} = \%10.33$$

$$MdRE = \%9.83$$

| StackExchange PostgreSQL Actual-Prediction (Monthly) (% Error) | | | |
|---|---|---|---|
| | Actual | Prediction | % Error |
| Jan-14 | 1104 | 963.56 | 12.72% |
| Feb-14 | 1215 | 973.39 | 19.89% |
| Mar-14 | 1403 | 983.23 | 29.92% |
| Apr-14 | 1241 | 993.07 | 19.98% |
| May-14 | 1249 | 1002.91 | 19.70% |
| Jun-14 | 1192 | 1012.75 | 15.04% |
| Jul-14 | 1298 | 1022.59 | 21.22% |
| Aug-14 | 1118 | 1032.43 | 7.65% |
| Sep-14 | 1269 | 1042.27 | 17.87% |
| Oct-14 | 1271 | 1052.10 | 17.22% |
| Nov-14 | 1278 | 1061.94 | 16.91% |
| Dec-14 | 1178 | 1071.78 | 9.02% |
| Jan-15 | 1321 | 1081.62 | 18.12% |
| Feb-15 | 1358 | 1091.46 | 19.63% |

***Table 6.67:** Monthly Relative error of PostgreSQL Prediction on StackExchange Data*

$$MRE = \frac{1}{N}\sum_{1}^{N}\frac{A_i - P_i}{A_i} = \%17.49$$

$$MdRE = \%17.99$$

## 2.7. MICROSOFT ACCESS

As being a long forgotten solution for single user desktop applications Access still performs well against its limitations. But this unfortunately this performance drops drastically ever year. In Tables 6.68 and 6.69 shows less and less usage time passes. Those numbers could easily interpreted as there are better option try those.

| StackExchange Access Actual-Prediction (Quarterly) (% Error) | | | |
|---|---|---|---|
| | Actual | Prediction | % Error |
| 2014 Q1 | 2339 | 2377.10 | 1.63% |
| 2014 Q2 | 2333 | 2576.79 | 10.45% |
| 2014 Q3 | 2085 | 2776.48 | 33.16% |
| 2014 Q4 | 1853 | 2976.17 | 60.61% |

***Table 6.68:*** *Quarterly Relative error of Microsoft Access Prediction on StackExchange Data*

$$MRE = \frac{1}{N}\sum_{1}^{N}\frac{A_i - P_i}{A_i} = \%26.46$$

$$MdRE = \%21.81$$

| StackExchange Access Actual-Prediction (Monthly) (% Error) | | | |
|---|---|---|---|
| | Actual | Prediction | % Error |
| Jan-14 | 709 | 703.73 | 0.74% |
| Feb-14 | 796 | 715.94 | 10.06% |
| Mar-14 | 834 | 728.14 | 12.69% |
| Apr-14 | 898 | 740.34 | 17.56% |
| May-14 | 719 | 752.55 | 4.67% |
| Jun-14 | 716 | 764.75 | 6.81% |
| Jul-14 | 799 | 776.95 | 2.76% |
| Aug-14 | 688 | 789.16 | 14.70% |
| Sep-14 | 598 | 801.36 | 34.01% |
| Oct-14 | 675 | 813.56 | 20.53% |
| Nov-14 | 610 | 825.77 | 35.37% |
| Dec-14 | 568 | 837.97 | 47.53% |
| Jan-15 | 722 | 850.17 | 17.75% |
| Feb-15 | 677 | 862.38 | 27.38% |

***Table 6.69:*** *Monthly Relative error of Microsoft Access Prediction on StackExchange Data*

$$MRE = \frac{1}{N}\sum_{1}^{N}\frac{A_i - P_i}{A_i} = \%18.04$$

$$MdRE = \%17.66$$

## 2.8. AMAZONRDS

Even the numbers are small Tables 6.70 and 6.71 shows a consistency in count of posts that mentions Amazon RDS. Cause of this stability is mostly belong to the stability of the Amazon Web Service and having invested in this platform is mostly related to the Cloud platform that you will invest.

| StackExchange Amazon RDS Actual-Prediction (Quarterly) (% Error) | | | |
|---|---|---|---|
| | Actual | Prediction | % Error |
| 2014 Q1 | 78 | 61.22 | 21.51% |
| 2014 Q2 | 66 | 64.99 | 1.53% |
| 2014 Q3 | 78 | 68.76 | 11.84% |
| 2014 Q4 | 63 | 72.53 | 15.13% |

***Table 6.70:*** *Quarterly Relative error of Amazon RDS Prediction on StackExchange Data*

$$MRE = \frac{1}{N}\sum_{1}^{N}\frac{A_i - P_i}{A_i} = \%12.50$$

$$MdRE = \%13.49$$

| StackExchange Amazon RDS Actual-Prediction (Monthly) (% Error) | | | |
|---|---|---|---|
| | Actual | Prediction | % Error |
| Jan-14 | 22 | 19.51 | 11.33% |
| Feb-14 | 24 | 19.89 | 17.14% |
| Mar-14 | 32 | 20.27 | 36.67% |
| Apr-14 | 25 | 20.65 | 17.42% |
| May-14 | 21 | 21.02 | 0.12% |
| Jun-14 | 20 | 21.40 | 7.02% |
| Jul-14 | 26 | 21.78 | 16.22% |
| Aug-14 | 23 | 22.16 | 3.64% |
| Sep-14 | 29 | 22.54 | 22.27% |
| Oct-14 | 21 | 22.92 | 9.14% |
| Nov-14 | 19 | 23.30 | 22.63% |
| Dec-14 | 23 | 23.68 | 2.95% |
| Jan-15 | 19 | 24.06 | 26.62% |
| Feb-15 | 35 | 24.44 | 30.18% |

***Table 6.71:*** *Monthly Relative error of Amazon RDS Prediction on StackExchange Data*

$$MRE = \frac{1}{N}\sum_{1}^{N}\frac{A_i - P_i}{A_i} = \%15.95$$

$$MdRE = \%16.82$$

# 3. CLOUD SERVICES

## 3.1. AMAZON WEB SERVICES

As expected Amazon Web Services has most impact in developer community. Tables 6.72 and 6.73 shows this impact increases constantly. Success of the Amazon Web Services model still continuous even with powerful competitors like Microsoft Azure. It is sure that success of this technology makes sure that it is a good technology that you may invest on it.

| Amazon Actual (Quarterly) (% Error) | | | |
|---|---|---|---|
| | Actual | Prediction | % Error |
| 2014 Q1 | 3602 | 2975.31 | 17.40% |
| 2014 Q2 | 3755 | 3215.54 | 14.37% |
| 2014 Q3 | 3981 | 3455.77 | 13.19% |
| 2014 Q4 | 4222 | 3696.00 | 12.46% |

***Table 6.72:*** *Quarterly Relative error of Amazon Web Services Prediction on StackExchange Data*

$$MRE = \frac{1}{N}\sum_{1}^{N}\frac{A_i - P_i}{A_i} = \%14.35$$

$$MdRE = \%12.72$$

| Amazon Actual (Monthly) (% Error) | | | |
|---|---|---|---|
| | Actual | Prediction | % Error |
| Jan-14 | 1073 | 940.97 | 12.30% |
| Feb-14 | 1139 | 948.85 | 16.69% |
| Mar-14 | 1390 | 956.72 | 31.17% |
| Apr-14 | 1347 | 964.60 | 28.39% |
| May-14 | 1223 | 972.48 | 20.48% |
| Jun-14 | 1185 | 980.35 | 17.27% |
| Jul-14 | 1357 | 988.23 | 27.18% |
| Aug-14 | 1234 | 996.10 | 19.28% |
| Sep-14 | 1390 | 1003.98 | 27.77% |
| Oct-14 | 1408 | 1011.85 | 28.14% |
| Nov-14 | 1415 | 1019.73 | 27.93% |
| Dec-14 | 1399 | 1027.61 | 26.55% |
| Jan-15 | 1449 | 1035.48 | 28.54% |
| Feb-15 | 1629 | 1043.36 | 35.95% |

***Table 6.73:*** *Monthly Relative error of Amazon Web Services Prediction on StackExchange Data*

$$MRE = \frac{1}{N}\sum_{1}^{N}\frac{A_i - P_i}{A_i} = \%24.83$$

$$MdRE = \%26.39$$

## 3.2. MICROSOFT AZURE

As popularity of cloud systems increase Azure's popularity increases too. And Tables 6.74 and 6.74 shows this increase is cannot be underestimated. Nearly more than %30 more than predicted value actual impact of Azure in developer community is on the rise. This could be mostly about free tier services and price drops but still a huge increase makes Azure assertive in cloud services sector.

| Azure Actual-Prediction (Quarterly) (% Error) | | | |
|---|---|---|---|
| | Actual | Prediction | % Error |
| 2014 Q1 | 2434 | 2003.64 | 17.68% |
| 2014 Q2 | 2931 | 1977.61 | 32.53% |
| 2014 Q3 | 3045 | 1951.91 | 35.90% |
| 2014 Q4 | 3024 | 1926.55 | 36.29% |

***Table 6.74:*** *Quarterly Relative error of Microsoft Azure Prediction on StackExchange Data*

$$MRE = \frac{1}{N}\sum_{1}^{N}\frac{A_i - P_i}{A_i} = \%30.60$$

$$MdRE = \%34.21$$

| Azure Actual-Prediction (Monthly) (% Error) | | | |
|---|---|---|---|
| | Actual | Prediction | % Error |
| Jan-14 | 734 | 645.96 | 11.99% |
| Feb-14 | 829 | 633.57 | 23.57% |
| Mar-14 | 871 | 621.41 | 28.66% |
| Apr-14 | 1007 | 609.48 | 39.48% |
| May-14 | 1030 | 597.79 | 41.96% |
| Jun-14 | 894 | 586.31 | 34.42% |
| Jul-14 | 988 | 575.06 | 41.80% |
| Aug-14 | 970 | 564.03 | 41.85% |
| Sep-14 | 1087 | 553.20 | 49.11% |
| Oct-14 | 1063 | 542.59 | 48.96% |
| Nov-14 | 1042 | 532.17 | 48.93% |
| Dec-14 | 919 | 521.96 | 43.20% |
| Jan-15 | 1114 | 511.95 | 54.04% |
| Feb-15 | 1256 | 502.12 | 60.02% |

***Table 6.75:*** *Monthly Relative error of Microsoft Azure Prediction on StackExchange Data*

$$MRE = \frac{1}{N}\sum_{1}^{N}\frac{A_i - P_i}{A_i} = \%40.57$$

$$MdRE = \%42.58$$

## 3.3. GOOGLE CLOUD PLATFORM

Tables 6.76 and 6.77 shows our forecast gives very similar values to the actual data. Even this is a good sign, while there are 2 big competitors available in cloud services sector it might not resulted good. This hesitant position is may be related to the Googles cloud services approach. Even they serve huge cloud services like Drive and Gmail their Cloud Platform is far behind of Amazon. Even they are closing the gap with new services they are still need to work harder.

| Google Actual (Quarterly) (% Error) | | | |
|---|---|---|---|
| | Actual | Prediction | % Error |
| 2014 Q1 | 2510 | 2340.41 | 6.76% |
| 2014 Q2 | 2865 | 2454.86 | 14.32% |
| 2014 Q3 | 2862 | 2569.32 | 10.23% |
| 2014 Q4 | 2567 | 2683.77 | 4.55% |

***Table 6.76:*** *Quarterly Relative error of Google Cloud Platform Prediction on StackExchange Data*

$$MRE = \frac{1}{N}\sum_{1}^{N}\frac{A_i - P_i}{A_i} = \%8.96$$

$$MdRE = \%8.49$$

| Google Actual (Monthly) (% Error) | | | |
|---|---|---|---|
| | Actual | Prediction | % Error |
| Jan-14 | 790 | 738.36 | 6.54% |
| Feb-14 | 758 | 750.32 | 1.01% |
| Mar-14 | 962 | 762.27 | 20.76% |
| Apr-14 | 1028 | 774.23 | 24.69% |
| May-14 | 961 | 786.19 | 18.19% |
| Jun-14 | 876 | 798.14 | 8.89% |
| Jul-14 | 1006 | 810.10 | 19.47% |
| Aug-14 | 953 | 822.06 | 13.74% |
| Sep-14 | 903 | 834.01 | 7.64% |
| Oct-14 | 847 | 845.97 | 0.12% |
| Nov-14 | 852 | 857.93 | 0.70% |
| Dec-14 | 868 | 869.88 | 0.22% |
| Jan-15 | 897 | 881.84 | 1.69% |
| Feb-15 | 890 | 893.79 | 0.43% |

***Table 6.77:*** *Monthly Relative error of Google Cloud Platform Prediction on StackExchange Data*

$$MRE = \frac{1}{N}\sum_{1}^{N}\frac{A_i - P_i}{A_i} = \%8.86$$

$$MdRE = \%8.26$$

## 3.4. HEROKU

Tables 6.78 and 6.79 shows Heroku is consistency with its small user base. Even there are small spikes present in their time frame this is mostly because of the already mentioned small user base. For a better insights about the future of Heroku we need more detailed research with additional data.

| Heroku Actual-Prediction (Quarterly) (% Error) | | | |
|---|---|---|---|
| | Actual | Prediction | % Error |
| 2014 Q1 | 1236 | 1185.06 | 4.12% |
| 2014 Q2 | 1187 | 1242.80 | 4.70% |
| 2014 Q3 | 1136 | 1295.79 | 14.07% |
| 2014 Q4 | 1024 | 1344.17 | 31.27% |

**Table 6.78:** Quarterly Relative error of Heroku Prediction on StackExchange Data

$$MRE = \frac{1}{N}\sum_{1}^{N}\frac{A_i - P_i}{A_i} = \%13.54$$
$$MdRE = \%9.38$$

| Heroku Actual-Prediction (Monthly) (% Error) | | | |
|---|---|---|---|
| | Actual | Prediction | % Error |
| Jan-14 | 404 | 372.91 | 7.70% |
| Feb-14 | 368 | 364.99 | 0.82% |
| Mar-14 | 464 | 357.06 | 23.05% |
| Apr-14 | 427 | 349.14 | 18.23% |
| May-14 | 409 | 341.22 | 16.57% |
| Jun-14 | 351 | 333.30 | 5.04% |
| Jul-14 | 398 | 325.38 | 18.25% |
| Aug-14 | 378 | 317.46 | 16.02% |
| Sep-14 | 360 | 309.53 | 14.02% |
| Oct-14 | 353 | 301.61 | 14.56% |
| Nov-14 | 333 | 293.69 | 11.80% |
| Dec-14 | 338 | 285.77 | 15.45% |
| Jan-15 | 431 | 277.85 | 35.53% |
| Feb-15 | 390 | 269.93 | 30.79% |

**Table 6.79:** Monthly Relative error of Heroku Prediction on StackExchange Data

$$MRE = \frac{1}{N}\sum_{1}^{N}\frac{A_i - P_i}{A_i} = \%16.27$$
$$MdRE = \%16.29$$

# 4. MOBILE OPERATING SYSTEMS

## 4.1. ANDROID

Table 6.80 and 6.81 shows that Android passed IOS not only in market share, also Developer community of Android more active than iOS community. Testing of forecast shows actual count of posts that mention Android increases faster than predicted data. In light of this data we may say Android will be continue its leadership in Mobile Operating Systems.

| StackExchange Android Actual-Prediction (Quarterly) (% Error) | | | |
|---|---|---|---|
| | Actual | Prediction | % Error |
| 2014 Q1 | 70186 | 56806.86 | 19.06% |
| 2014 Q2 | 67659 | 54825.19 | 18.97% |
| 2014 Q3 | 72992 | 52912.65 | 27.51% |
| 2014 Q4 | 68611 | 51066.83 | 25.57% |

**Table 6.80:** Quarterly Relative error of Android Prediction on StackExchange Data

$$MRE = \frac{1}{N}\sum_{1}^{N}\frac{A_i - P_i}{A_i} = \%22.78$$

$$MdRE = \%22.32$$

| StackExchange Android Actual-Prediction (Monthly) (% Error) | | | |
|---|---|---|---|
| | Actual | Prediction | % Error |
| Jan-14 | 22378 | 19786.35 | 11.58% |
| Feb-14 | 22344 | 19911.62 | 10.89% |
| Mar-14 | 25464 | 20034.37 | 21.32% |
| Apr-14 | 23155 | 20154.67 | 12.96% |
| May-14 | 21530 | 20272.57 | 5.84% |
| Jun-14 | 22974 | 20388.10 | 11.26% |
| Jul-14 | 26362 | 20501.33 | 22.23% |
| Aug-14 | 23991 | 20612.29 | 14.08% |
| Sep-14 | 22639 | 20721.03 | 8.47% |
| Oct-14 | 22535 | 20827.60 | 7.58% |
| Nov-14 | 22982 | 20932.04 | 8.92% |
| Dec-14 | 23094 | 21034.39 | 8.92% |
| Jan-15 | 23528 | 21134.69 | 10.17% |
| Feb-15 | 23034 | 21232.98 | 7.82% |

***Table 6.81:*** *Monthly Relative error of Android Prediction on StackExchange Data*

$$MRE = \frac{1}{N}\sum_{1}^{N}\frac{A_i - P_i}{A_i} = \%11.57$$

$$MdRE = \%9.55$$

## 4.2. IOS

Tables 6.82 and 6.83 shows mention count of IOS not as much as expected in predictions. In detail an increasing trend seems turned into stationary trend. A detailed research with additional data could give some answers about this turn in the trend.

| **StackExchange IOS Actual-Prediction (Quarterly) (% Error)** | | | |
|---|---|---|---|
| | Actual | Prediction | % Error |
| 2014 Q1 | 28595 | 30996.79 | 8.40% |
| 2014 Q2 | 27463 | 35102.74 | 27.82% |
| 2014 Q3 | 32105 | 39752.59 | 23.82% |
| 2014 Q4 | 30818 | 45018.37 | 46.08% |

***Table 6.82:*** *Quarterly Relative error of IOS Prediction on StackExchange Data*

$$MRE = \frac{1}{N}\sum_{1}^{N}\frac{A_i - P_i}{A_i} = \%26.53$$

$$MdRE = \%25.82$$

| StackExchange IOS Actual-Prediction (Monthly) (% Error) | | | |
|---|---|---|---|
| | Actual | Prediction | % Error |
| Jan-14 | 9167 | 8434.91 | 7.99% |
| Feb-14 | 9161 | 8772.12 | 4.24% |
| Mar-14 | 10267 | 9111.20 | 11.26% |
| Apr-14 | 9251 | 9451.74 | 2.17% |
| May-14 | 8626 | 9793.31 | 13.53% |
| Jun-14 | 9586 | 10135.53 | 5.73% |
| Jul-14 | 10405 | 10478.00 | 0.70% |
| Aug-14 | 9687 | 10820.35 | 11.70% |
| Sep-14 | 12013 | 11162.22 | 7.08% |
| Oct-14 | 11848 | 11503.26 | 2.91% |
| Nov-14 | 10126 | 11843.12 | 16.96% |
| Dec-14 | 8844 | 12181.50 | 37.74% |
| Jan-15 | 9431 | 12518.07 | 32.73% |
| Feb-15 | 9705 | 12852.56 | 32.43% |

***Table 6.83:*** *Monthly Relative error of IOS Prediction on StackExchange Data*

$$MRE = \frac{1}{N}\sum_{1}^{N}\frac{A_i - P_i}{A_i} = \%13.37$$

$$MdRE = \%11.48$$

## 4.3. WINDOWS PHONE

Similar to Android Tables 6.84 and 6.85 shows count of posts that mentions Windows Phone is not as much as expected in predictions. Also a detailed research is needed for give some answers about this turn in the trend.

| StackExchange Windows Phone Actual-Prediction (Quarterly) (% Error) | | | |
|---|---|---|---|
| | Actual | Prediction | % Error |
| 2014 Q1 | 4097 | 3983.25 | 2.78% |
| 2014 Q2 | 4184 | 4186.23 | 0.05% |
| 2014 Q3 | 3809 | 4384.04 | 15.10% |
| 2014 Q4 | 3011 | 4576.82 | 52.00% |

***Table 6.84:*** *Quarterly Relative error of Windows Phone Prediction on StackExchange Data*

$$MRE = \frac{1}{N}\sum_{1}^{N}\frac{A_i - P_i}{A_i} = \%17.48$$

$$MdRE = \%8.94$$

| StackExchange Windows Phone Actual-Prediction (Monthly) (% Error) | | | |
|---|---|---|---|
| | Actual | Prediction | % Error |
| Jan-14 | 1353 | 1221.00 | 9.76% |
| Feb-14 | 1323 | 1352.99 | 2.27% |
| Mar-14 | 1421 | 1323.00 | 6.90% |
| Apr-14 | 1450 | 1420.99 | 2.00% |
| May-14 | 1408 | 1450.00 | 2.98% |
| Jun-14 | 1326 | 1408.00 | 6.18% |
| Jul-14 | 1362 | 1326.01 | 2.64% |
| Aug-14 | 1282 | 1362.00 | 6.24% |
| Sep-14 | 1165 | 1282.01 | 10.04% |
| Oct-14 | 1075 | 1165.01 | 8.37% |
| Nov-14 | 1010 | 1075.01 | 6.44% |
| Dec-14 | 926 | 1010.01 | 9.07% |
| Jan-15 | 989 | 926.01 | 6.37% |
| Feb-15 | 996 | 988.99 | 0.70% |

**Table 6.85:** Monthly Relative error of Windows Phone Prediction on StackExchange Data

$$MRE = \frac{1}{N}\sum_{1}^{N}\frac{A_i - P_i}{A_i} = \%5.71$$

$$MdRE = \%6.30$$

# 5. INTERPRETATIONS

| Knowledge Area | Technologies | Monthly Mean Relative Error | Monthly Median Relative Error | Quarterly Mean Relative Error | Quarterly Median Relative Error |
|---|---|---|---|---|---|
| Programming Languages | JavaScript | 9.00% | 7.18% | 10.80% | 9.68% |
| | Java | 15.02% | 14.16% | 9.11% | 9.29% |
| | PHP | 13.63% | 13.86% | 12.73% | 12.55% |
| | C# | 21.52% | 24.83% | 9.03% | 7.73% |
| | Python | 10.68% | 12.23% | 11.69% | 10.96% |
| | C++ | 21.07% | 26.78% | 15.52% | 14.97% |
| | SQL | 25.51% | 17.52% | 24.44% | 11.69% |
| | C | 22.34% | 30.25% | 19.15% | 21.45% |
| | R | 12.80% | 13.93% | 8.09% | 7.38% |
| | Objective C | 19.15% | 15.02% | 8.96% | 10.13% |
| | VB.NET | 18.67% | 21.01% | 12.49% | 13.22% |
| | Matlab | 12.87% | 13.75% | 14.36% | 12.94% |
| | Scala | 6.78% | 7.47% | 2.92% | 2.99% |
| | Perl | 32.05% | 34.47% | 22.69% | 20.50% |
| | Haskell | 11.49% | 11.22% | 9.67% | 9.52% |
| | Go | 15.91% | 14.98% | 28.96% | 29.62% |
| | Assembly | 19.28% | 16.62% | 23.24% | 21.86% |
| | XSLT | 31.38% | 35.51% | 20.79% | 21.69% |
| | Ruby | 8.06% | 8.27% | 15.06% | 10.73% |
| Databases | MS SQL Server | 10.45% | 11.66% | 9.05% | 8.46% |
| | MySQL | 16.69% | 20.32% | 10.98% | 12.41% |
| | Oracle | 7.34% | 6.47% | 5.61% | 6.06% |
| | MongoDB | 41.33% | 47.04% | 23.49% | 23.90% |
| | DB2 | 8.60% | 4.92% | 6.70% | 5.07% |
| | PostgreSQL | 17.49% | 17.99% | 10.33% | 9.83% |
| | AmazonRDS | 15.95% | 16.82% | 12.50% | 13.49% |
| Mobile OS | Android | 11.57% | 9.55% | 22.78% | 22.32% |
| | IOS | 13.37% | 11.48% | 26.53% | 25.82% |
| | Windows Phone | 5.71% | 6.30% | 17.48% | 8.94% |

**Table 6.86:** General Table for Mean and Medians of Relative Errors

As we seen from the relative error values, in general forecast library of R is performs really well. Except some unexpected result like Table 6.85, which possible caused by having not enough data; test results gave errors in acceptable ranges. As we seen in the Table 6.86 forecasts with high percentage error rates are old or new technologies which are already on decreasing or increasing usage trend.

Especially forecasts of short time ranges are exceedingly good. As a matter of fact if we analysis long range forecasts (more than 6 month) we can see that those long range forecasts increases the Relative Error in our predictions.

As a result we can recommend usage of this library in forecasting applications.

# 7 CONCLUSIONS AND RECOMMENDATIONS

Results of the forecasts gives us some insights about the trends of the Information Technologies. Some of promising technologies are really performed well with respect to their forecast in tests. For example R is really performed well against our tests. R, performs at least 10 percent better than forecasted in every prediction point.

Haskell is another promising language that performs better than expected than forecasts. As we taken account the performance of the Scala and R, we can say that functional languages get into programming scheme more and more.

As another newcomer language Go performs exceedingly well. As Google's support is cannot be ignored we can see Go more projects in the future.

In Database Technology knowledge area only rising RDBM System is PostgreSQL. We think this rise is related to their industry standard implementations like PostGIS and its scaling ability.

MongoDB as a document based database performs pretty good against test data. It performs up to 50 percent than predicted. As we taken account the rise of the NoSQL database that is not a surprise.

In Cloud Services knowledge area even Amazon Web Services continues its reign in top of the market it still rises in means of mentioned posts. It performs better than predictions and we think this is the result of being innovative.

Microsoft Azure continues its chase of Amazon Web services. It performs 30 to 60 percent better than forecasted values. It seems Microsoft is determined about catching Amazon in Cloud Services market.

In Mobile Operating Systems knowledge area Android comes from behind and passed IOS for a few year for now. And its velocity still increasing as it is only Mobile OS performs better than predictions. Drop in IOS is really conspicuous. As it really performs poor against forecasts especially after 3rd quarter of 2014. Perhaps this information could be a good starting point for a future research.

If we make some recommendations for future researches.

In smaller size researches (smaller set of Knowledge Areas like Technologies) we can add "Users" to dimensionality of the data and using relations between technologies and users we can create coefficients for analysis.

Analysis of graph metrics of our data is could be a start point of a future research.

As there is not enough data present in GitHub we can extract data from "Event" messages and use those messages for a thru research. But processing those messages could require a serious processing power.

Significant drop of count of posts which mentions SQL in 4th quarter of 2014 very steep this could be a research topic.

With spread of Swift a comparative research with Objective C could give interesting results.

# APPENDIX A
# STACKEXCHANGE DATA STRUCTURE

| Posts |
| --- |
| Id |
| PostTypeId |
| AcceptedAnswerId |
| ParentID |
| CreationDate |
| Score |
| ViewCount |
| Body |
| OwnerUserId |
| OwnerDisplayName |
| LastEditorUserId |
| LastEditorDisplayName |
| LastEditDate |
| LastActivityDate |
| Title |
| Tags |
| AnswerCount |
| CommentCount |
| FavoriteCount |
| ClosedDate |
| CommunityOwnedDate |

| Users |
| --- |
| Id |
| Reputation |
| CreationDate |
| DisplayName |
| LastAccessDate |
| WebsiteUrl |
| Location |
| AboutMe |
| Views |
| UpVotes |
| DownVotes |
| EmailHash |
| AccountId |
| Age |

s

| Comments |
|---|
| Id |
| PostId |
| Score |
| Text |
| CreationDate |
| UserDisplayName |
| UserId |

| Badges |
|---|
| Id |
| UserId |
| Name |
| Date |
| Class |
| TagBased |

| CloseAsOffTopicReasonTypes |
|---|
| Id |
| IsUniversal |
| MarkdownMini |
| CreationDate |
| CreationModeratorId |
| ApprovalDate |
| ApprovalModeratorId |
| DeactivationDate |
| DeactivationModeratorId |

| PendingFlags |
|---|
| Id |
| FlagTypeId |
| PostId |
| CreationDate |
| CloseReasonTypeId |
| CloseAsOffTopicReasonTypeId |
| DuplicateOfQuestionId |
| BelongsOnBaseHostAddress |

| PostFeedback |
|---|
| Id |
| PostId |
| IsAnonymous |
| VoteTypeId |
| CreationDate |
| PostHistory |

| PostHistory |
|---|
| Id |
| PostHistoryTypeId |
| PostId |
| RevisionGUID |
| CreationDate |
| UserId |
| UserDisplayName |
| Comment |
| Text |

| PostLinks |
|---|
| Id |
| CreationDate |
| PostId |
| RelatedPostId |
| LinkTypeId |

| PostTags |
|---|
| PostId |
| TagId |

| ReviewRejectionReasons |
|---|
| Id |
| Name |
| Description |
| PostTypeId |

| ReviewTaskResults |
| --- |
| Id |
| ReviewTaskId |
| ReviewTaskResultTypeId |
| CreationDate |
| RejectionReasonId |
| Comment |

| ReviewTasks |
| --- |
| Id |
| ReviewTaskTypeId |
| CreationDate |
| DeletionDate |
| ReviewTaskStateId |
| PostId |
| SuggestedEditId |
| CompletedByReviewTaskId |

| SuggestedEdits |
| --- |
| Id |
| PostId |
| CreationDate |
| ApprovalDate |
| RejectionDate |
| OwnerUserId |
| Comment |
| Text |
| Title |
| Tags |
| RevisionGUID |

| SuggestedEditVotes |
| --- |
| Id |
| SuggestedEditId |
| UserId |
| VoteTypeId |
| CreationDate |
| TargetUserId |
| TargetRepChange |

| Tags |
| --- |
| Id |
| TagName |
| Count |
| ExcerptPostId |
| WikiPostId |

| TagSynonyms |
| --- |
| Id |
| SourceTagName |
| TargetTagName |
| CreationDate |
| OwnerUserId |
| AutoRenameCount |
| LastAutoRename |
| Score |
| ApprovedByUserId |
| ApprovalDate |

| Votes |
| --- |
| Id |
| PostId |
| VoteTypeId |
| UserId |
| CreationDate |
| BountyAmount |

# APPENDIX B
# GITHUB EVENTS

| |
|---|
| CommitCommentEvent |
| CreateEvent |
| DeleteEvent |
| DeploymentEvent |
| DeploymentStatusEvent |
| DownloadEvent |
| FollowEvent |
| ForkEvent |
| ForkApplyEvent |
| GistEvent |
| GollumEvent |
| IssueCommentEvent |
| IssuesEvent |
| MemberEvent |
| MembershipEvent |
| PageBuildEvent |
| PublicEvent |
| PullRequestEvent |
| PullRequestReviewCommentEvent |
| PushEvent |
| ReleaseEvent |
| RepositoryEvent |
| StatusEvent |
| TeamAddEvent |
| WatchEvent |